\documentclass[manuscript]{acmart}

\let\citep\cite
\let\citet\cite

\AtBeginDocument{%
  \providecommand\BibTeX{{%
    \normalfont B\kern-0.5em{\scshape i\kern-0.25em b}\kern-0.8em\TeX}}}

\usepackage{graphics}      

\def\plaintitle{Hierarchical Decision Ensemble(s)- An inferential framework for uncertain Human-AI collaboration in forensic examinations}

\def\emptyauthor{}
\def\plainkeywords{Forensics; Firearm Examination; Machine Learning; Statistical Models; Inferential framework; Calibrating Trust; Human-AI collaboration}

\definecolor{linkColor}{RGB}{6,125,233}
\hypersetup{%
  pdftitle={\plaintitle},
  pdfauthor={\emptyauthor},
  pdfkeywords={\plainkeywords},
  pdfdisplaydoctitle=true, 
  bookmarksnumbered,
  pdfstartview={FitH},
  colorlinks,
  citecolor=black,
  filecolor=black,
  linkcolor=black,
  urlcolor=linkColor,
  breaklinks=true,
  hypertexnames=false
}

\begin{document}

\title{\plaintitle}


%

    \author{Ganesh Krishnan}
    \affiliation{
     \institution{Iowa State University}
     \streetaddress{}
     \city{Ames}
     \state{Iowa}
     \country{US}}
    \email{ganeshk@iastate.edu}
    \author{Heike Hofmann}
    \affiliation{
     \institution{Iowa State University}
     \streetaddress{}
     \city{Ames}
     \state{Iowa}
     \country{US}}
    \email{hofmann@iastate.edu}

%

\begin{abstract}
Forensic examination of evidence like firearms and toolmarks, traditionally involves a visual and therefore subjective assessment of similarity of two questioned items. Statistical models are used to overcome this subjectivity and allow specification of error rates. These models are generally quite complex and produce abstract results at different levels of the analysis. Presenting such metrics and complicated results to examiners is challenging, as examiners generally do not have substantial statistical training to accurately interpret results. This creates distrust in statistical modelling and lowers the rate of acceptance of more objective measures that the discipline at large is striving for. We present an inferential framework for assessing the model and its output. The framework is designed to calibrate trust in forensic experts by bridging the gap between domain specific knowledge and predictive model results, allowing forensic examiners to validate the claims of the predictive model while critically assessing results.
\end{abstract}


\keywords{\plainkeywords}

%

\maketitle

\hypertarget{introduction}{%
\section*{Introduction}\label{introduction}}
\addcontentsline{toc}{section}{Introduction}

Statistical modeling, which is not restricted to machine learning, is a key part of all AI systems. Even the simplest AI system, where the automation is restricted to specific activities, involves complex underlying models that produces results at many different levels. These statistical and machine learning models, almost always produce results that range from: raw data, parameter estimates, feature or covariate maps, covariate distributions, statistics such as goodness-of-fit for a model, scores, bounds, intervals etc. When statisticians and model experts present results to domain experts, who usually have a limited understanding of statistical modeling techniques, difficulty arises in explaining the model results and laying out their meaning in a way that makes them useful, and understandable for the domain experts. However, understanding modeling and results are fundamentally important for better decision making. In this paper, we present an inferential framework dealing with this situation in the field of forensic examination, where the need is dire, and the means and methods to address such problems, minimal. To do this, we first reveal the true nature of the problem in the next few sections, and show by example both, the AI side and the human side of decision-making, before synthesizing an inferential framework and presenting its design for Human-AI collaboration. After which we argue, about validity of such systems in calibrating trust and augmenting human decision-making under uncertainty. We rely on the fact, that designing effective collaborative systems for Human-AI interfacing, especially in high-stakes fields like forensics, call for deep considerations of both the AI side of the things and the examiners side of what it means to make a correct decision. Our goal in presenting this work, is therefore, to not only show how we can design Examiner-AI collaboration in forensics, but to also present the immediate need for such systems, as the field of forensics transitions from manual procedures to AI augmented decision-making. Thus, we showcase a prevalent problem in a critical field, that has not been addressed yet, to a wider audience. In doing so, we argue, that designing systems that are fundamentally and theoretically sound is paramount in such critical fields and leads to efficient communication of uncertainties. We also show that, to do so we cannot just rely on recommendations, explanations and interpretations, as they do not necessarily mean inference drawn from such systems can be considered accurate. Hence, Human-AI collaborative systems in fields with high-stakes, need a good base and robust framework for addressing accurate calibration of trust.

\hypertarget{visual-comparison-in-forensics}{%
\section*{Visual Comparison in Forensics}\label{visual-comparison-in-forensics}}
\addcontentsline{toc}{section}{Visual Comparison in Forensics}

Forensic examination of evidence, as part of a criminal investigation, involves comparison of different pieces of evidence and determining whether these pieces come from the same source or from a different source. Many sub-fields of forensic evidence examination, like firearms \citep{riva_comparison_2020}\citep{afte-toolmarks1998}, toolmarks \citep{mattijssen_firearm_2021}\citep{afte-toolmarks1998}, facial image comparison \citep{white_perceptual_2015}\citep{growns_human_2020}, fingerprint comparison \citep{dror_contextual_2006}\citep{growns_human_2020}, document examination \citep{dyer_visual_2006}\citep{growns_human_2020} etc have a significant reliance on visual comparison of the evidence by forensic examiners.

The subjective nature of these examinations and the lack of error rates associated with this process was first criticized by the National Academy of Sciences in 2009 \citep{NAS:2009} and re-iterated by the President's Council of Advisors on Science and Technology \citep{PCAST} in 2016. In response, the discipline has been striving to move towards statistical modeling procedures that bring objectivity to the analysis. Objectivity achieved through predictive modeling often means dealing with complex methods, algorithms and abstract results at different stages of an analysis. Conveying these results and procedures to forensic examiners is challenging as examiners typically do not have an in-depth statistical training or expertise in statistical modeling. The nature of the field means that any decisions made by examiners affect criminal investigations and propagate to judicial proceedings, thus having widespread implications. Therefore, the ability to accurately interpret modeling results and draw correct conclusions, is of paramount importance. With this in mind, we want to address how to bridge the gap between domain specific knowledge of the forensic examiners and statistical model results.

Thus, our objective in this paper, is to present a framework, that demonstrates a balance between calibrating trust in the model while simultaneously encouraging a critical assessment of the statistical results. The framework promotes this balance by ensuring that there is no over-reliance on numbers, and the cognitive load is kept low, so that the examiners can focus on drawing correct conclusions as their trust in the results gets calibrated.

In this work, we take the example of forensic firearm examination as a procedural proxy for visual comparisons of forensic evidence in general. We explain by example the requirements, problems, concerns and possible solutions to them, which would be applicable to the visual comparisons of most forensic evidence and not just firearms. Currently firearm examiners, visually compare markings found on the surface of fired bullets (or cartridge casings), under a comparison microscope, and conclude whether the markings came from the same gun or not. A similar process is also true for visual comparison of other evidence, where the difference would be in the choice of specific features.

\begin{figure}
\includegraphics[width=\columnwidth]{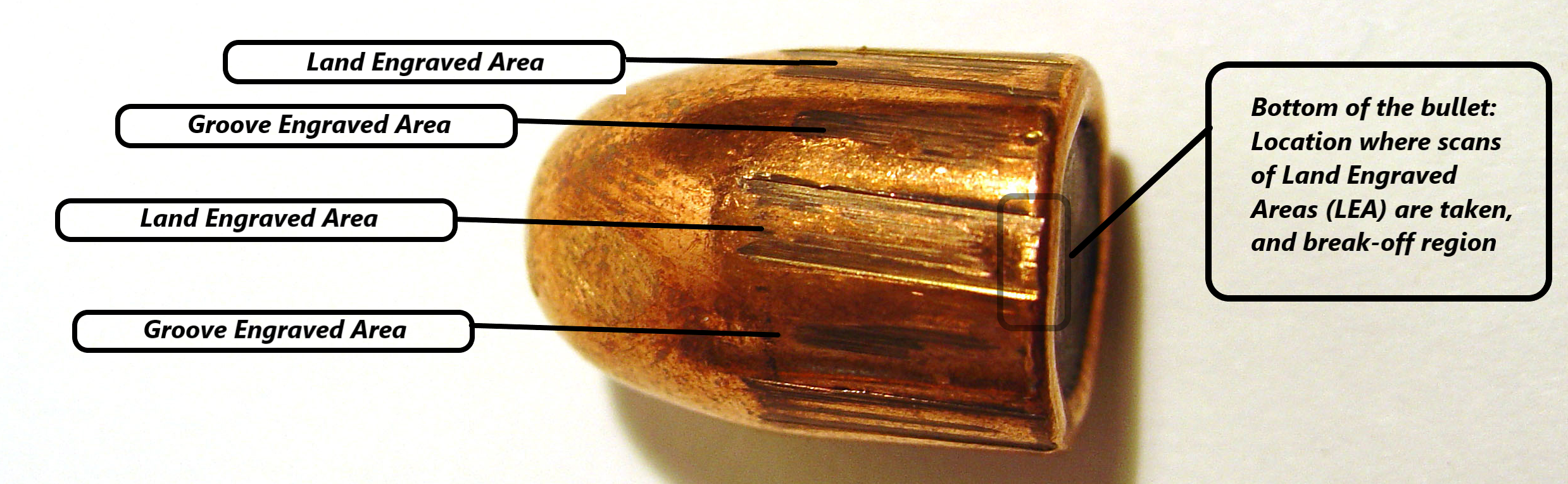}
\caption{\label{comparison-nist} Bullet collected after it was fired from a gun. The LEAs, GEAs and the location where LEA scans are taken can be seen here. Unannotated source: NIST}
\label{fig:bullet_gea_lea}
\end{figure}

\hypertarget{complexity-in-design-of-human-ai-interfaces}{%
\section*{Complexity in Design of Human-AI Interfaces}\label{complexity-in-design-of-human-ai-interfaces}}
\addcontentsline{toc}{section}{Complexity in Design of Human-AI Interfaces}

The complexity of designing Human-AI interfaces has been discussed by
Yang et al.~\citep{yang_re-examining_2020} in their recent work on. They discuss the factors that contribute to the complexity of such interfaces and show that, the success of a Human-AI system designs largely depend on uncertainty around the AI's capabilities and the output complexity of the AI results. They go on to reveal the challenges in development of such systems lies in the availability of AI related things that can be implemented in the design, crafting thoughtful interactions, and collaborating with AI engineers, but their discussion was mostly from the perspective of AI designers with AI engineers in the loop.

For forensic settings, the problems revealed by Yang et al.~\citep{yang_re-examining_2020} gets even more complicated, because domain experts play a vital role. Domain experts in forensics, are the primary human agents involved in the decision-making process, but their knowledge base is considered significantly different from what is desired for optimal Human-AI collaboration. As such, it is also not feasible to expect deep probabilistic reasoning and complex quantitative rationale from the forensic experts that is equivalent to that of a model expert or statistician.

Hong et al\citep{hong_human_2020} explain this in their work by pointing towards the increasing scholarly interest in model interpretability in research communities like ML and HCI, but lack of any knowledge about what practitioners perceive and understand from the presented interpretability. The emphasize that without the understanding of what is being practiced, the research on interpretability would fall short on addressing critical needs, which could in turn lead to unrealistic solutions.

The complexity of designing Human-AI systems in forensics, is therefore, a bigger challenge, that many other fields are also likely to face. Before attempting to model, design or address this problem, it is important to first reveal the true nature of the problem, that we propose to address.

The underlying challenge for designing a robust and effective Human-AI system in forensic examination stems from the fact that the decision-making responsibility lies with the domain expert and not the model expert. Therefore, accurate inference under uncertainty is a major issue. Expanding the knowledge base and training, to achieve reliable collaborative decision-making with AI remain far-reaching goals of the community and achieving Human-AI collaboration by addressing, just these goals are not feasible, at least in the short run. The transitioning nature of the field means algorithmic aversion \citep{dietvorst_2015_algorithm} remains a strong contributing factor to acceptance of new modeling methods.

Facing these complex issues, we propose a Human-AI framework for high-stakes fields like forensic examinations, that bridge the gap between the statistical inference and perceptual inference. To evaluate the possibility of bridging this gap, the work of Buja et al.~\citep{buja_2009_statistical}
provides evidence to having one such kind of visual inference framework capable of utilizing the human cognition equivalent to formal statistical testing. Therefore, here, we leverage well established aspects of human cognition, by designing a Human-AI framework that serves as a function that maps the key perceptual inference mechanism, to the modeling steps of the underlying AI and hence, model inference.

Another, important point is the need of matched inferential frameworks as opposed to just explainable and transparent AI frameworks. Inferential frameworks, are more than just explainable and interpretable systems. They serve as protocol for congruent mapping of cognitive models with AI models and hence incorporate explanations within them as long as they follow the protocol. Explanations, on the other hand, do not address an important point, that existence of explanations and their provisions, does not necessarily correspond to coherent and accurate conclusions. When the human agent interacting with an explanation system is unaware of how the explanations assimilate, and lead to the final results given by the AI, the conclusions drawn are not necessarily accurate. This becomes complicated when uncertainty around the AI results is significant.

Zhang et al.~\citep{zhang_how_nodate} show that opaqueness and complexity are not the core issues in Human-AI interaction, but uncertainty is, and as such, focusing on explainable and transparent AI for high-stakes situations does not solve the key issues that arise from uncertainty in the AI results.

On a similar note, Dodge et al.\citep{dodge_after_2021} show that explainable AI systems although gaining traction and importance in modern society, have not been evaluated for their potential in assessing an AI. They present model-free and model-based explanations, for sequential decision-making frameworks, argue about better AI assessment when the human agents were augmented by their established protocol and action review system, a system that provided a specified process on when to rely on the AI results.

Gu et al.\citep{gu_lessons_2021} also reveal the challenges in bridging the gap between pathologists and AI, and suggest collaborative techniques around the AI's limitations as much as its capabilities.

Calibration of trust, is therefore a consequence of, congruent inferential frameworks where perceptual inference and statistical inference have a well defined correspondence, and this leads to matched conceptualization of uncertainty. Calibration of trust, established in this manner, should lead to faithful conclusions being drawn from the model even under uncertainty. One way to address congruence in inferential process, is to leverage procedural overlaps that exist in the visual comparison process and AI comparison process.

To elicit the importance of calibrated trust in forensics, on a slightly different note, Growns et al.~\citep{growns_human_2020} talk about the need for accurate conceptualization of underlying cognitive mechanisms in visual comparisons. They also talk about augmentation of purely visual comparison procedures with statistical metrics as cues to improve forensic visual comparison, but they fall short of discussing or considering collaborative decision-making where AI/model results are to be inferred, interpreted and communicated by the forensic examiners. Kelman et al.~\citep{kellman_forensic_2014} on the other hand noted that more complicated the image,the more difficult the manual comparison procedure and the less confident examiners are in their decisions, but without being able to explain why.

As such, collaborative systems are yet to be adopted and accepted in forensics, even when the underlying AI has superior performance and quantification of uncertainties. The adoption of AI based results, evaluation of the error rates and uncertainties around it, and the adaptability of the entire field of forensics to the new age of robust and objective results, thus relies on the accurate calibration of trust elicited earlier.

\hypertarget{bullet-matching-in-objective-and-subjective-settings}{%
\section*{Bullet matching in objective and subjective settings}\label{bullet-matching-in-objective-and-subjective-settings}}
\addcontentsline{toc}{section}{Bullet matching in objective and subjective settings}

\begin{figure}
\includegraphics[width=0.7\textwidth]{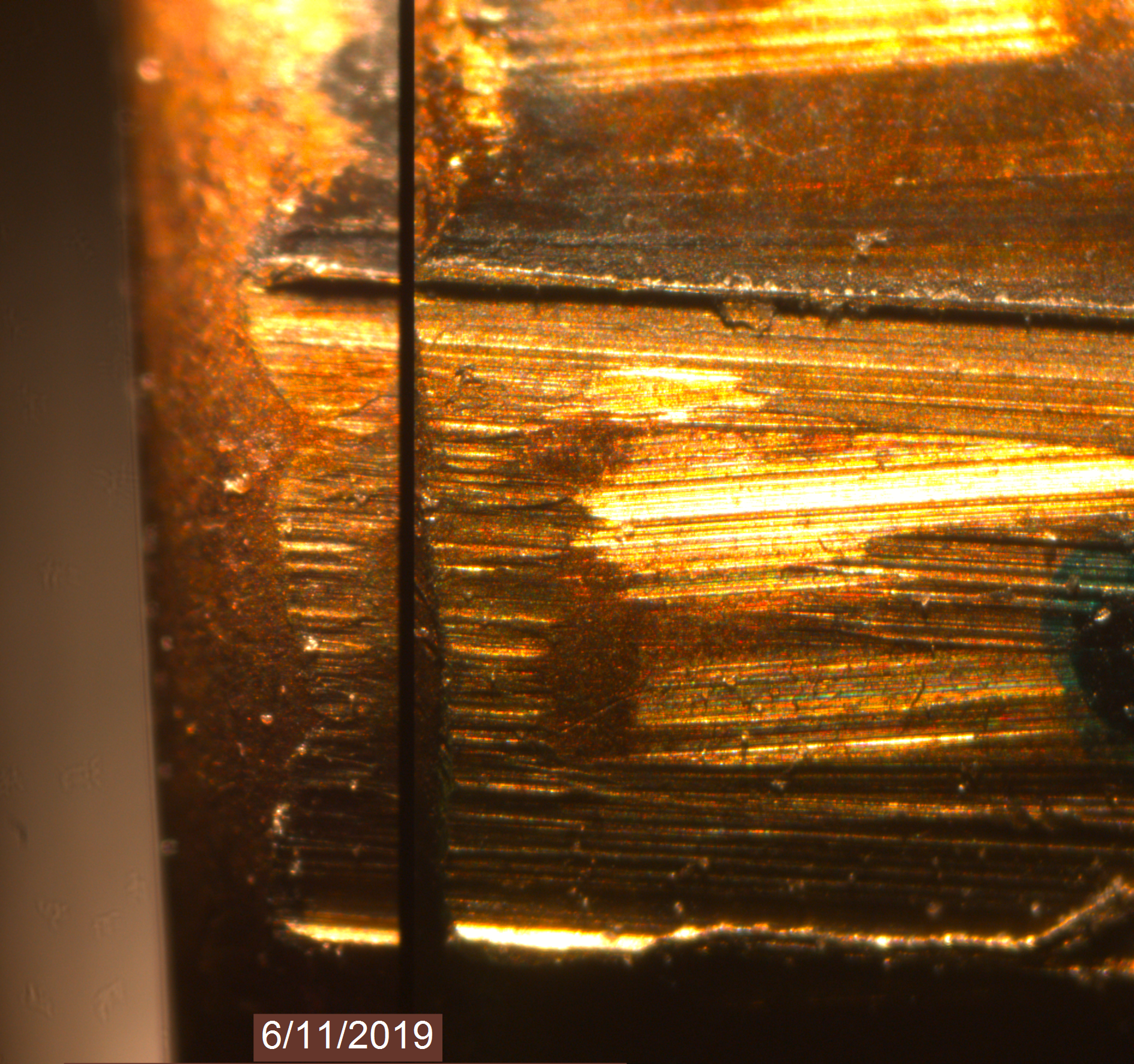}
\caption{\label{comparison1}Image depicts what firearms examiners see in a comparison microscope. These are two bullets aligned one over the other. The black vertical line separates between two bullets. The striation marks are visible on both the LEAs and the firearm examiners tries to align them by rotating one of the bullets till they seem to align correctly. If two LEAs are aligned correctly, the examiner visually makes a comparison and deduction on whether the two LEAs seem to match. If they conclude its a match, they quickly make synchronized in-phase rotations of both bullets to compare the other 5 pairs too.}
\label{fig:visualcomparison}
\end{figure}

Markings on the surface of a bullet are made by the rifling action of the gun from which it is fired. These markings are striated and have been suggested by the forensic community to be a means for uniquely identifying the barrel of the gun from which it is fired \citep{afte-article1992}. The striation marks are located on specific regions on the surface of the bullet called land engraved areas (LEAs). Depending on the firearm, there are many LEAs in a bullet and each LEA is separated by a groove engraved areas (GEA). Figure \ref{fig:bullet_gea_lea} shows where these LEAs and GEAs are located on the surface of a bullet. Owing to the contour of the bullet, which slowly tapers down from the bottom or heel to the nose of the bullet, the most prominent markings are found near the base of the bullet. This is because the cross-sectional area on the LEAs near to the base, has the maximum contact with barrel of the gun. Therefore, both firearm examiners and automatic methods alike, use striations which are near the bottom of the bullet. In case the heel of the bullet is damaged, a region that is undisturbed by the damage and yet closest to the base, is considered to be the best possible region within the given LEA, as this is where prominent striation marks are expected to be found. To make a comparison firearm examiners, start with the alignment of bullet LEAs. The circular cross-section of a bullet can have a few LEAs and GEAs. The number usually hovers around 5 or 6 but can also go upto 8 LEAs per bullet. This number is a characteristic of the firearm and thus, depends on the lands and grooves found in the barrel of the gun. Given the quasi-cylindrical surface contour of the bullet, LEAs on each bullet can be imagined to be poised in a circular fashion forming the cylinder. This means if the scans of LEAs are taken separately, there is only one sequence of circular arrangement of the LEA scans that depicts the bullet correctly. A matching pair of bullet will therefore, usually have only one sequence of pairs of LEAs that have maximum agreement. This means, for the case that there are 6 LEAs on the two bullets being questioned, if we start by naming the LEAs on one bullet as B1-1 to B1-6, and the LEAs on the second bullet as B2-1 to B2-6, there will exist only one sequence of consecutively positioned LEAs on bullet 1 which will be in maximum agreement with only one sequence of consecutively positioned LEAs on bullet 2. This is usually not a problem for firearm examiners as they deal with physical bullets where the sequence remains a part of the form of the bullet. The firearm examiner start their comparison by first positioning the bullets over one another under a comparison microscope. When the positioning is stable, they match these striation marks \citep{afte-article1992} by making visual comparisons between them. Figure \ref{fig:visualcomparison} shows what firearms examiner see when making such a visual comparison under a comparison microscope. The firearm examiners then assess whether the markings from the two bullets are in ``sufficient agreement'' or not \citep{afte-toolmarks1998} by accounting for the striated nature of the surface of the bullet LEAs in a non-quantitative manner. Since the examiners start by considering striae on specific LEAs of the two bullets, a vital step in their visual comparison procedure is to ensure the sequence of the two bullets are in-phase. This means when looking at two physical bullets, that are apriori known to be matches, the examiner just has to make sure they find the right starting pair of LEAs amongst either (B1-1, B2-1) or (B1-1, B2-2) and so on till (B1-1, B2-6), and see if they have the location of the grooves and most prominent striations aligned. The firearm examiners do so by visually confirming the alignment on the starting pair and then rotate through the two bullets, while maintaining the phase difference in the rotation of the two bullets. This means 5 other LEA pairing combinations between the two bullets are checked to confirm that they all can be aligned in a manner that lets the examiners conclude, through their visual comparison that they are matches. At times, due to randomness of the firing process or the bullet hitting the target sideways, some lands can be damaged or may not have as much visual agreement as the examiner would like it to be. But nevertheless, for known matches, all 6 pairs of LEAs are expected to be matching. In case of non-matching bullets, on the other hand, this becomes a harder task and takes a cognitive toll on the examiner as they have to now go through and check all combinations that might lead to a sequence of matching pairs.

A usual step in assistance to the visual comparison task employed by the examiners, is to use some kind of counting or scaling metric to support their claims. One such metric used by firearm examiners is the counting of striae peaks that match. These peaks are minuscule protrusions while valleys being minuscule dents on the surface that make the pattern look striated. The examiners simply count the peaks of the striations that have been visually confirmed to match between two LEAs, and see how many such peaks match consecutively. The valleys between two striations marks are harder to visually identify as the comparison microscope cannot capture the light in these very small fissures between two peaks.

These counting or scaling metrics serve merely as cues and small checkpoints that are used to support the mental claim of the examiner whether the two questioned LEAs that they visually compared were matching or not-matching. But these so called scaling or counting of peaks metrics cannot be used on their own to adjudge two comparisons as matches as there is no evidence to suggest that they can stand alone as classification metrics. This is primarily because what constitutes as matching peaks or matching valleys remains a subjective decision by the examiner, making the whole exercise subjective. The cognitive rationale for visually matching peaks is therefore non-quantifiable in nature. This means that the cognitive rationale for matching peaks remains latent and may or may not be the same as any other metric that has been derived mathematically to identify the peaks and valleys in the LEA striations. On the surface it may seem like, the counting of peaks done by the examiners, for assisting the visual comparison, is some kind of a numeric quantity, hence making the endeavor objective in nature. But in reality, owing to the non-quantifiable cognitive rational behind the choice of peaks, the count is subjective to what the examiner considers as matching striae. Apart from this aided count, the visual comparison process involves other non-trivial cognitive maneuvers of pattern recognition that are hard to explain. Therefore, the visual comparison procedure remains non-exploitable and intractable as an objective measure. Each comparison, depends on the firearm examiner conducting the investigation and the conclusions can vary significantly between examiners. The non-quantifiable nature of this setting means, any claims or conclusions made, remain highly susceptible to bias. Figure \ref{fig:visualcomparison} shows how forensic examiner position two bullets in a comparison microscope and try make this visual comparison to identify if the observed LEAs match or not.

\begin{figure}
\includegraphics[width=\columnwidth]{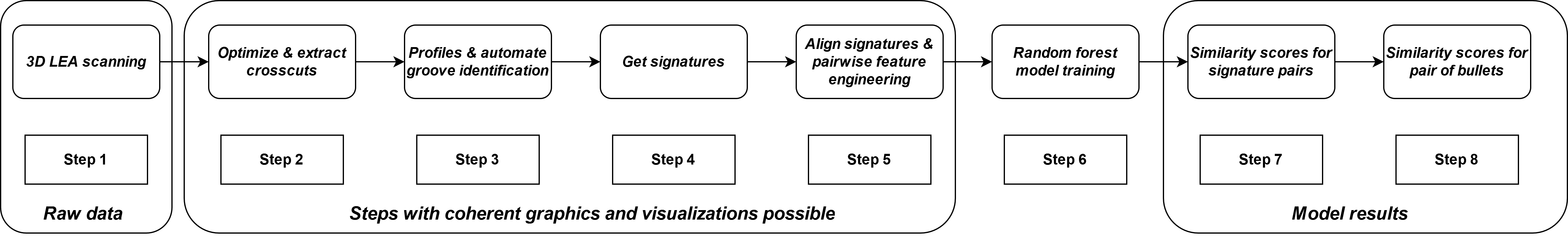}
\caption{\label{comparison-pipeline} The different steps in the bullet matching pipeline for automatic matching of bullets is presented here. The data flow starts with the scanning process where the LEAs are scanned, before proceeding to model development and information extraction, and then model results which include pairwise land-to-land scores, and bullet level scores.}
\label{fig:bullet-matching-pipeline}
\end{figure}

Objective measures on the other hand try to quantify different features of a bullet LEA and use them in statistical methods to come up with scores. There have been a few such measures developed in the past for comparing two bullets with the purpose of identifying their source \citep{aoas}\citep{ma2004}\citep{gkhh}\citep{chu2013}. Hare et al.~\citep{aoas} proposed a random forest based machine learning method for comparing two bullets. In their work they show a standard procedure of moving from data to model to random forest scores to bullet scores. Krishnan and Hofmann \citep{gkhh} later employed a similar pipeline to come up with a non-parametric score of comparison. The standard steps of this bullet matching pipeline can be seen in the Figure \ref{fig:bullet-matching-pipeline}. The raw data used in this process is in the form of 2D or 3D scanned images of the bullet LEAs. The process starts by identifying the LEAs and GEAs in the bullet scans. The LEAs have striation marks on their surface, which is needed for making a comparison. Depending on the category of the firearm, there can be different number of LEAs and GEAs that can be seen on bullet surfaces. To avoid confusion created by multiple firearms and bullets, and for the sake of explanations, we will consider a specific class of firearm, the Ruger pistols. This allows us to keep the number of LEAs and GEAs constant in the discussions. The bullets fired through these Ruger pistols have 6 LEAs on it. The next step in the bullet matching pipeline is to find a suitable region to extract markings that can be used in statistical methods. A cross-sectional area near the base of the bullet is supposed to have the most prominent set of markings. The typical process of extracting the most suitable cross-sectional marking is an automatic computation which involves calculating the cross-correlation between different markings from the same LEA. Other complicated steps of identification of missing data are also a part of this process. The computations of this step therefore, identifies a region free from irregularities like break-off etc and a marking that is most representative of the global LEA structure. Thus, for each of the 6 LEAs a suitable cross-section location is chosen and the marking corresponding to this cross-cut is extracted. We call this extracted marking a `profile' \citep{gkhh}\citep{aoas}. Profiles reflect the curvature and other specific traits of an LEA. After this, grooves are identified on the profiles using some automated identification methods. After extracting the profile and removing the grooves, a smoothing function called the LOWESS fit, \citep{lowess} which is similar to a gaussian filter, is applied on the profile to remove the curvature and any unnecessary noise from the data. The residuals of the lowess fit are called `signatures'. These signatures reflect all the characteristics that are required for an objective comparison method to come up with a score.

From a cognitive stand point, signatures are representative abstractions of the striated markings present on the LEA and therefore, give a different level of understanding of the intricacies of the surface of the LEA. Profiles are similarly important, and reflect the large scale structure of the LEA across a cross-section. They give a comparative visual summary and relative understanding of the small scale structure of the striations when compared to the large scale structure like curvature of the bullet LEA. Both these visual abstractions namely, signatures and profiles are not used by the examiner, but connect the firearm examiners understanding of the physical bullet to the structure that essentially drives objective assessments. There is a lot of powerful inference that can be drawn by looking and contemplating on these visualizations, which cannot be otherwise drawn by simply looking at the bullet structure. Therefore such abstractions and derivations of the image, can be used to connect and bridge the scores generated by statistical models to the raw data that is the scans of the LEAs.

\begin{figure}
\includegraphics[width=.99\textwidth]{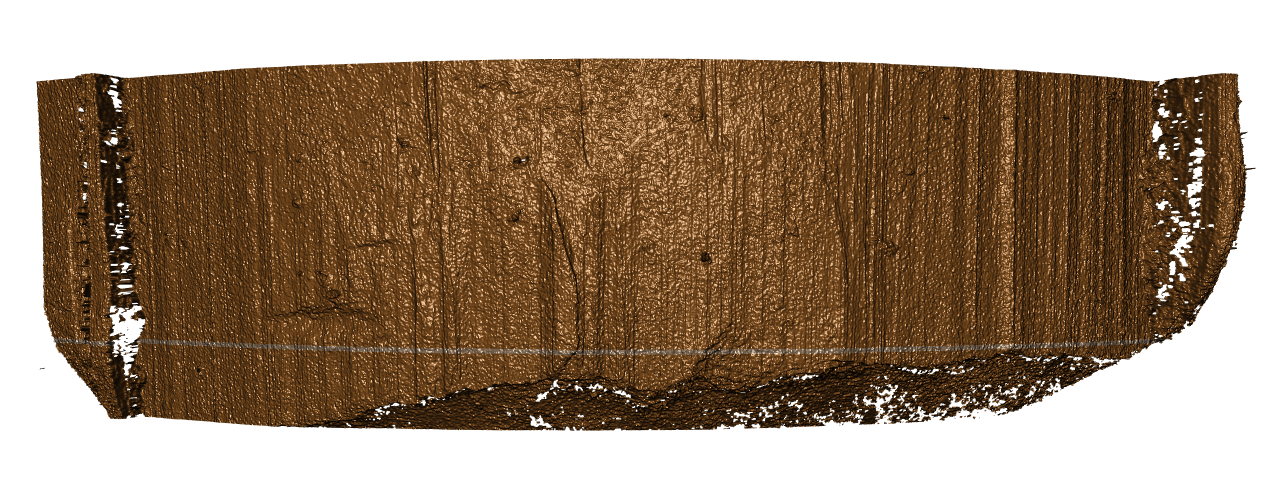}
\caption{The image shows a rendering of the scan of the bullet land engraved area (LEA) which includes the grooves on either side and a break-off region at the bottom. The break-off region shows the flaking off of parts of the bullet, which happens after the firing process of the gun. The process initiates when the firing pin hits the bottom, and the breaking off at the bottom can happen any time between then and the bullet moving through the barrel of the gun and hitting its target. The light silver-grey line shows the region from which cross-cuts are extracted for automatic matching of the bullet lands. \label{fig:x3p-image}}
\end{figure}

\begin{figure}
\includegraphics[width=.8\textwidth]{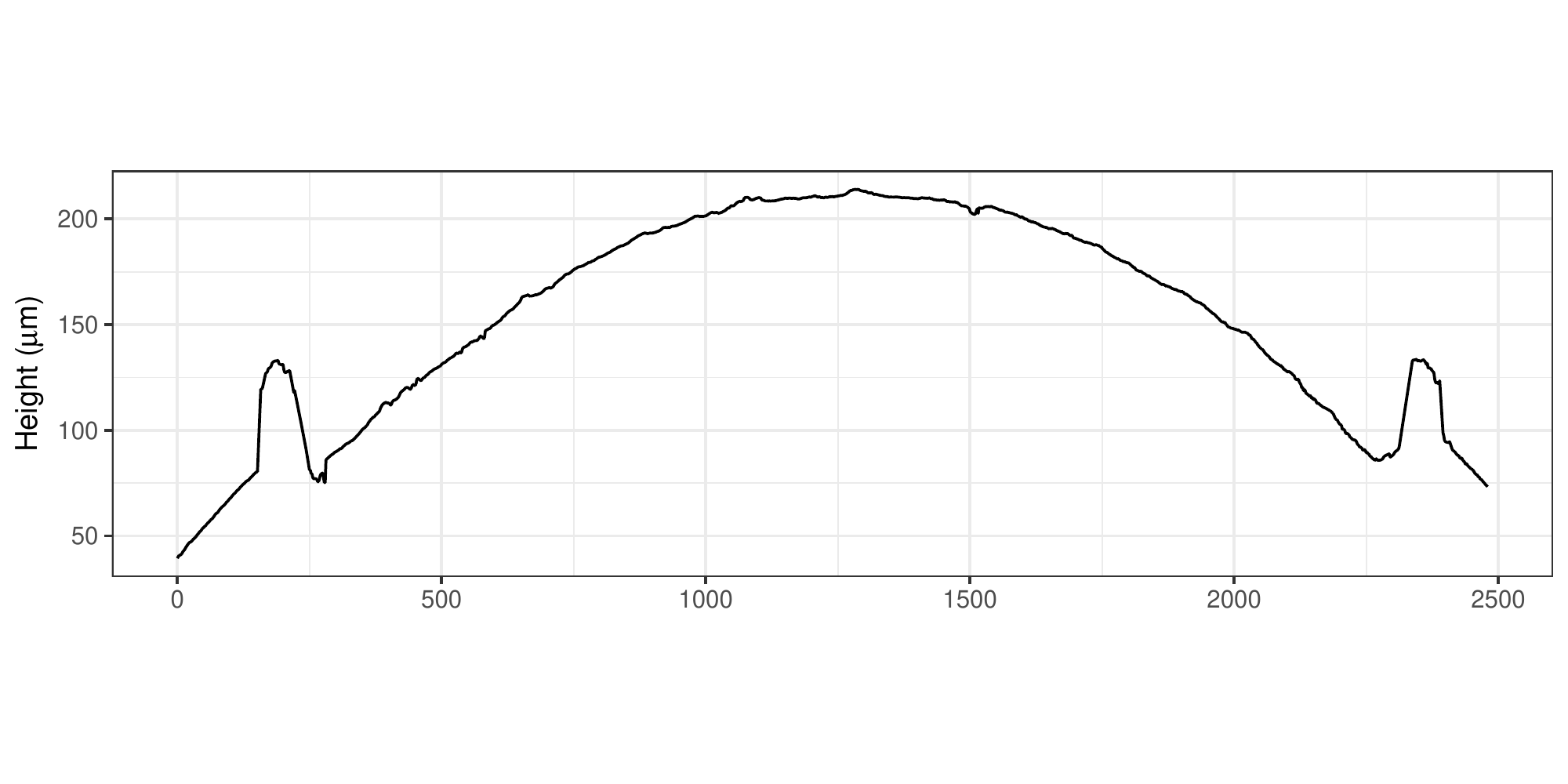} \caption{The crosscut values shown are extracted at a fixed distance from the bottom of the land. These profiles therefore represent the scrapings on the land surface but at a fixed height from the bottom.}\label{fig:x3p-profile}
\end{figure}

\begin{figure}
\includegraphics[width=.8\textwidth]{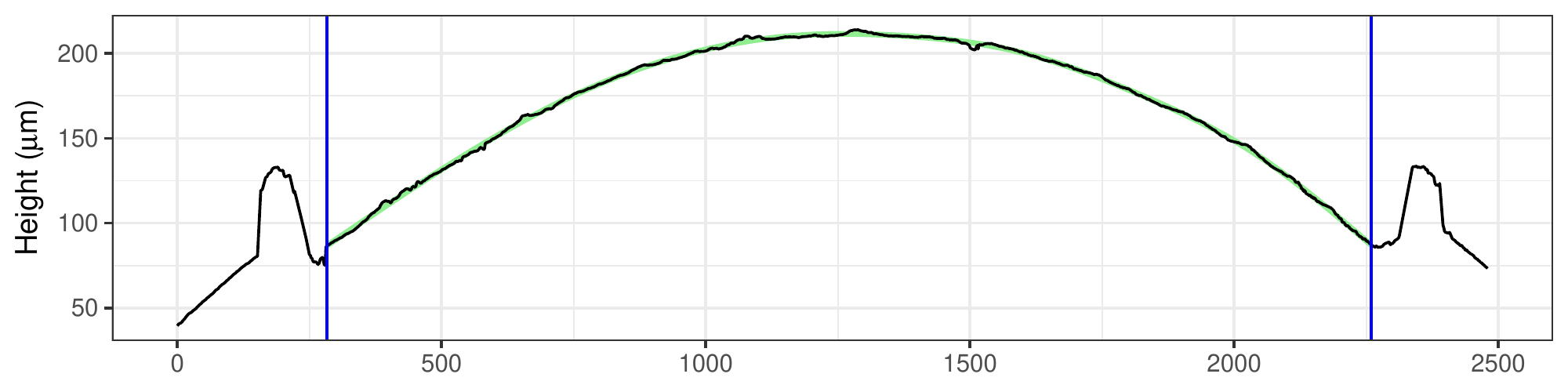} \caption{The image shows grooves marked in blue on the profiles. The identification of grooves on the LEA scan is done using an automatic groove identification algorithm. The part of the profile between the two grooves is used for analysis. The green part shows a smoothed marking that is used in signatures.}\label{fig:x3p-grooves}
\end{figure}

\begin{figure}
\includegraphics[width=.8\textwidth]{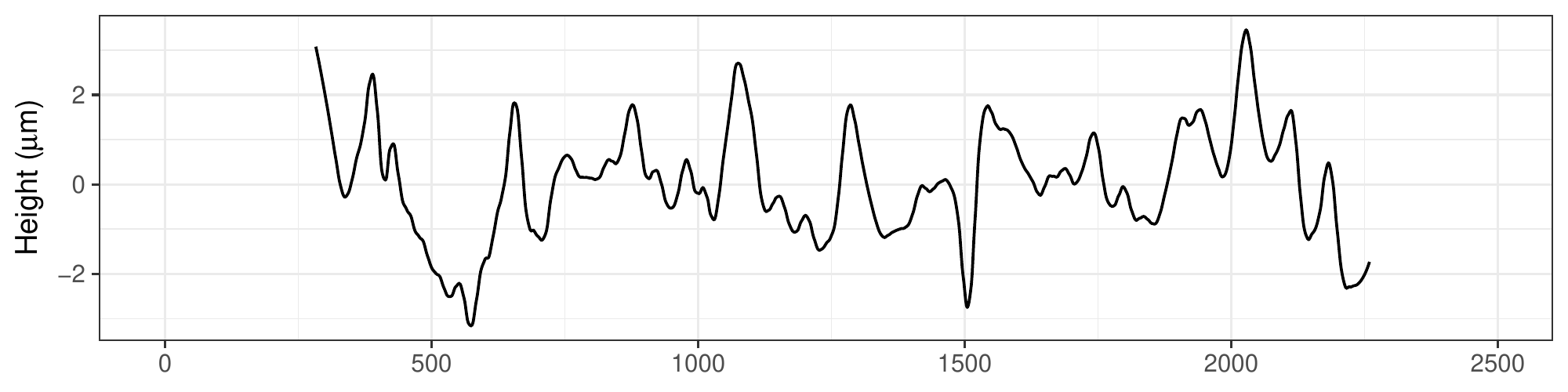} \caption{The marking shown here is the signature of the land which is extracted by further processing the profiles by removing the trend of the cross-section but still preserving the striations observed on the land surface. These land signatures are used for comparing two lands}\label{fig:x3p-signature}
\end{figure}

\begin{figure}
\includegraphics[width=.8\textwidth]{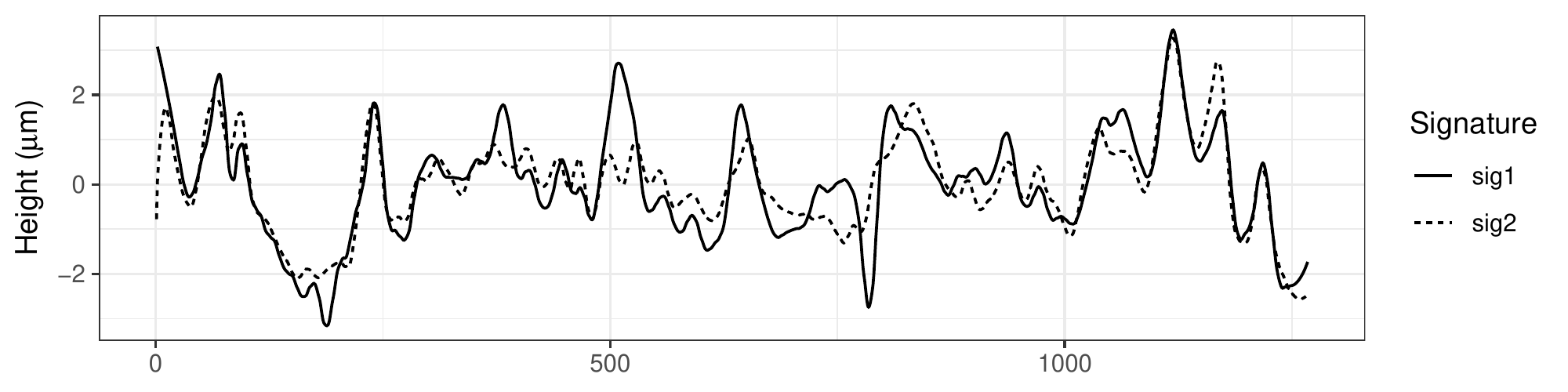} \caption{The algorithmic alignment of two signatures extracted are shown here. The markings are aligned by using the regions of maximum agreement which is computed by maximizing the cross-correlation between them}\label{fig:x3p-align}
\end{figure}

\begin{figure}
\includegraphics[width=.8\textwidth]{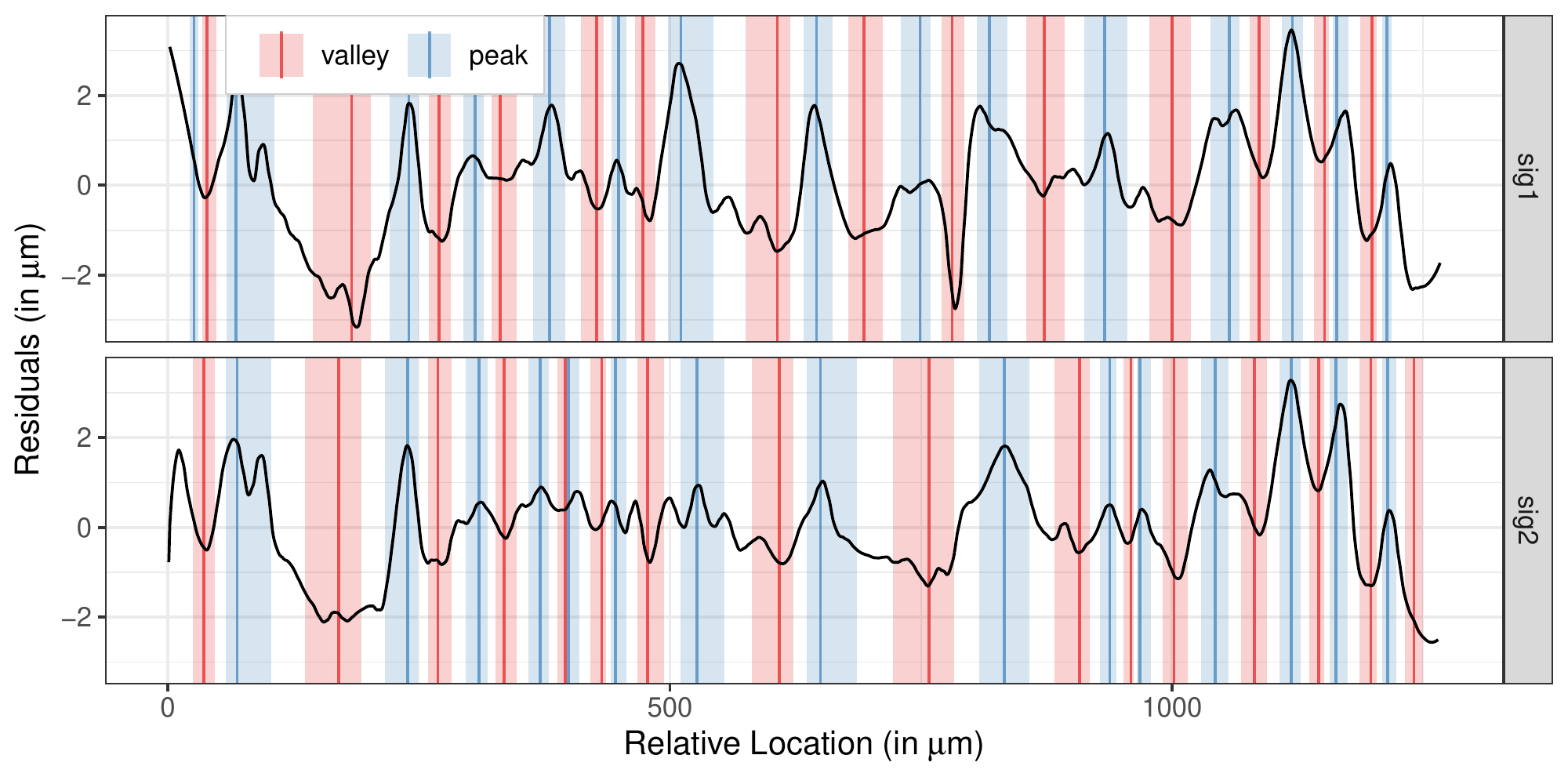} \caption{The figure depicts automatically calculated peaks (blue) and valleys (red), and their width based on signatures extracted from a land engraved area. Two different signatures can be visually compared to a certain extent based on how well the peaks and valleys seem to align. The automatic calculation metric is based on Consecutively Matching Striae (CMS) computations where the identification of these peaks, valleys and their widths use a well-quantified procedure. This is different from the visual comparison made by examiners where the peaks are counted based on intuitive recognition and identification}\label{fig:x3p-striation}
\end{figure}

The figure \ref{fig:x3p-profile} shows the cross-sectional profile after it is extracted from the LEA. The profile shows the two bumps on either side which represent the grooves. The figure \ref{fig:x3p-grooves} shows where the groove engraved areas (GEAs) are located by a groove detection algorithm for this particular profile. The detected groove in this case seem to be accurately identified. This identification serves an important purpose in the bullet matching pipeline, as GEAs are considered to be characteristics of the scan that hinder algorithmic matching. The correct identification of GEAs on the profiles allow the extraction of the useful part of the striation markings from the middle. After chopping of the grooves, the middle part of the profile, which is depicted in green in the figure \ref{fig:x3p-grooves}, is further processed to get the signatures. Figure \ref{fig:x3p-signature} shows one such extracted signature. It can be clearly seen that the removal of the large scale trend ensure small scale features can be viewed distinctly. Figure \ref{fig:x3p-align} shows two signatures aligned using the cross-correlation function and based on identification of regions of maximum agreement. This process is not the same as the alignment of striations done by firearm examiners. Figure \ref{fig:x3p-striation} shows two signatures where the peaks, valleys and the width of these extremas on the signatures have been identified using well quantified statistical metric. The automatically identified peaks and valleys show a similarity to the peak identification done by forensic examiners, but is not quite the same. The striation peaks identified by forensic examiners, are done visually and instinctively and their judgment is purely based on cognitive assessment. This is also true because forensic examiners do not work with digital data but use comparison microscopes where such computations are infeasible. Another issue, here is there is no conclusive proof that counting of peaks and other cognitive rationalizations of pattern recognition hold true always, because for one examiner this can be the same or very different from another examiner. The metric consecutively matching striae (CMS) \citep{chu2013} \citep{aoas}, that is closest to this, on the other hand, uses some underlying distance based calculations for identifying the peaks and valleys. Even then, when compared to other objective methods like statistical and machine learning models, the CMS on its own does not perform even close enough to the other methods to be advocated for use \citep{aoas} alone in the matching process. Therefore, it is not always prudent to use CMS as the only way to explain model results, but still, it can be viewed as a level of abstraction being offered about the different features of the signatures being compared. Combined with the right kind of visualizations and other metrics, it can direct the examiners to form a kind of rationale connecting the scores to different levels of these visual as well as numeric abstractions and then to the raw data i.e.~LEA scans.

\begin{figure}
\includegraphics[width=0.5\textwidth]{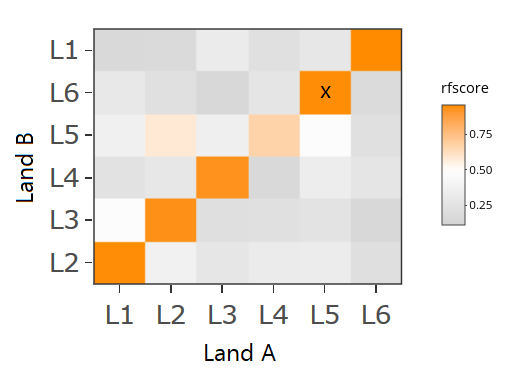}
\caption{\label{comparison-l2l} The land-to-land random forest score for 36 comparisons is shown here as a heatmap. The diagonal shows 6 pairs of in-phase comparisons. In-phase here means that, for two bullets fired from the same barrel, each of the 6 LEAs from one bullet will only have one other LEA on the second bullet to which it is supposed to match. Therefore, only 6 pairs from the 36 pair-wise comparisons will have higher match scores than the rest. This is an immediate extension of the firing phenomenon of the gun, where a barrel has 6 Lands that engrave 6 LEAs on the bullet, therefore, for two bullets fired from the same gun, each bullet will only have 1 LEA that passes through a specific land of the barrel. Thus, there can only be 6 pairs of LEAs owing to the 6 lands of the barrel, and these 6 pairs are supposed to be represented on the diagonal of the given heatmap.}
\label{fig:land_to_land}
\end{figure}

For objective methods, the signatures are data that is used and computations made on these signatures are used for comparisons. Machine learning models like the random forest, was first prescribed as a measure of comparison of bullet striation marks by Hare et al \citep{aoas}. The bullet matching pipeline has several complex steps of image processing already in place. To add to this, the random forest model takes many other statistical metrics like consecutively matching striae (CMS)\citep{chu2013}, cross-correlation function (CCF)\citep{vorburger2011}, average distance (D), number of matches and non-matches, sum of average heights of matched extrema etc \citep{aoas} as inputs. Without going into the details about these metrics, its important to first acknowledge the existence of such metrics, which on their own merit give some level of quantification by abstraction of signal information. It is also important to understand that with many such metrics being used in the random forest, the complexity of information that goes into machine learning algorithm is high. Moreover, given the inherent black-box nature of machine learning algorithms makes it harder to make within the model interpretations that can subsequently be used to bridge the gap between model results and the understanding that non-model experts and domain experts have. This is true for most statistical models which have complex sub-step procedures. In the context of matching bullets, the random forest based method proposed by Hare et al \citep{aoas} has been shown to give the best error rates, better than other metrics of comparison. Thus making the use of these machine learning models, the most likely objective measure of comparison for firearms. In this paper, we present a framework that tries to bridge the gap between modeling results and domain knowledge of the firearm examiners which is why our framework is presented primarily in the context of these random forest and machine learning scores, but most definitely holds for any statistical model that is performing a similar task

\begin{figure*}[htbp]
\includegraphics[width= 0.5\textwidth]{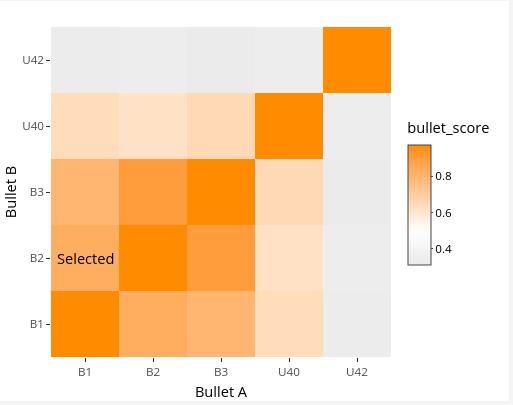} \caption{Heatmap showing the scores of 5 different bullets compared with each other making 25 different combinations. The diagonal of this heatmap shows the score for a bullet when compared with itself which will alway be 1. The rest of the comparisons show the statistical score for two different bullets that are being compared. These bullet-to-bullet scores incorporate the random forest scores of all in-phase pairs of land-to-land matches for two bullets under consideration. Uncertainty is embedded in the heatmap through the presence of multiple and similar such scores, due relative inference of the score of the comparison under consideration, with the scores of all other comparisons. There is also some ambiguity introduced due to the scales and color gradients}\label{fig:bullet_to_bullet}
\end{figure*}

\hypertarget{reasoning-in-forensic-examination}{%
\section*{Reasoning in forensic examination}\label{reasoning-in-forensic-examination}}
\addcontentsline{toc}{section}{Reasoning in forensic examination}

In the current setting of subjective comparisons, toolmark and firearm examiners, base their decisions on visual assessment of two questioned items under consideration. The current standards dictate that decisions made are of the form (a) identification, which means yes, the two objects come from the same source, (b) elimination meaning no, they do not come from the same source, (c) inconclusive evidence in data to support either yes or no and (d) data is unsuitable for decision making \citep{afte-toolmarks1998}.

Considering the already subjective nature of the analysis, the decision of inconclusive evidence is particularly a gray area in the field. In the absence of clear quantifiable cut-offs, what constitutes good evidence, to support yes or no, becomes a decision that is very susceptible to bias. At best, it relies on consensus, instead of hard facts and thresholds. This is a problem, as the current analysis practices of visual inspection, are already subjective. Therefore, in this case, it is not always simple to just provide heuristics as cues which the examiners can rely on. Nor is it prudent to push an examiner to make a decision which may turn out to be an incorrect one. What is needed is a critical assessment in an objective setting, so that each and every comparison made by an examiner is carefully investigated by going through a series of metrics, statistics and information which are products of an objective analysis. This minimizes the risk of making an incorrect decision, and at the same time enables the examiner to thoroughly examine the comparison. It also reduces the rate of ``inconclusive evidence in data to support yes or no''. Objective measures of comparison have another advantage. They provide a platform to move away from subjective binary decisions and towards drawing inference based on a critical analysis of various stages of results.

This is especially true when quantitative thresholds are absent, or when the scores are far away from these thresholds, and outside bounds that let us make binary classifications. Existence of such cases presents a challenge in making decisions, and therefore calls for a setup that allows for at least some level of inference to be drawn even in these uncertain scenarios.

Even in the absence of quantitative thresholds, objective statistical setups are better poised to address the challenge of drawing inference under uncertain conditions.
Objective analysis usually give probabilistic results with associated error rates and uncertainty information at different levels of the analysis. Each level of analysis has some kind of metric or statistical measure coming out. This gives the metrics a one-to-one correspondence with levels of the analysis. The information derived from the statistical results from these levels, eventually serves as data stages for a cognitive and psychological decision-making strategy which every person has. Binary decisions can only be considered when the model results of comparison are overwhelmingly close to a 0 or 1 in a probabilistic sense, or if thresholds are defined for such classification to be possible. Whether or not we have binary decisions, the uncertainty around the scores and what it means in the context of the domain, is necessary to be expressed in some way or the other. Therefore, the decisions are more than mere classifications and should be seen as a means for providing coherent and clear understanding of what the probabilistic result from the model stands for in the context of the raw-data at hand. The reason for using the context of raw data is because firearm examiners only deal with physical bullets in the current subjective analysis setting. Raw data which is 3D information of the surface of the bullet in the form of an images, can therefore be considered as instantiations of the physical bullet.

So, the decisions have to be in a form that takes the domain expertise of the firearm examiners and gives them ability to communicate the statistical results accurately in court proceedings. It is to be noted, that making domain experts statisticians, is not the goal, but it is possible to make them better in interpreting model results within the context of their domain, which is firearm examinations.

Therefore now the task is to make the examiners understand what different kinds of uncertainty and results means in the context of firearm examinations. Having this ability allows the domain expert to communicate the results as accurately as model experts but within the context of firearm examinations. It brings clarity and coherence to arguments made from objective comparison of two questionable items. Binary or not, results should always be considered with their associated uncertainty, inferred correctly without subjective bias or lack of model expertise affecting it, and communicated within the context of forensic examinations.

We, therefore, propose an interactive interface framework that calibrates trust in the model, enables critical thinking and keeps cognitive load at a minimum. The goal of the framework would be to enable the firearm examiners at each information step, make logical cognitive decision-strategies by enabling critical thinking, that lets them make coherent decisions on inference, and thereby communicate results in an unbiased way.

\hypertarget{hierarchical-decision-ensemble--an-inferential-framework-from-cognitive-theory}{%
\section*{Hierarchical decision ensemble- an inferential framework from cognitive theory}\label{hierarchical-decision-ensemble--an-inferential-framework-from-cognitive-theory}}
\addcontentsline{toc}{section}{Hierarchical decision ensemble- an inferential framework from cognitive theory}

The motivation for providing an interactive framework for calibrating trust and promoting critical reasoning in forensic firearm examination, comes from the evidence provided in cognitive theory. We follow principles that are grounded in user studies and use it as the foundation to propose our framework. We begin with a discussion of inherent decision making strategies in the presence of multiple statistics and probabilistic information. Meder and Mayrhofer \citep{meder_sequential} in their work, conduct an experiment where multiple statistics and other information sources exist, each of which, we refer to as a stage of information. In each information stage, both numerical results and approximate verbal cues derived from the corresponding numerics, is provided. These information stages were provided in a sequential setting. The authors then try to find whether decisions made by humans differ in any way, when only verbal cues are being considered in this sequential setting. In order to check this, on one hand, they had a statistical model that used the sequentially provided numerical values from the information stages to make a predictive decision. On the other hand, they had humans using the same sequential setting, but with only verbal cues to make a decision. Their goal was to dwell into diagnostic reasoning for humans when only approximate information is given sequentially. Meder et al\citep{meder_sequential} show that when the subjects were asked to make diagnostic reasoning leading to a final judgement, they had remarkably close correspondence with what a statistical model (as used by Meder et al.) \citep{meder_sequential} would predict. The sequential nature of the information being provided ensured, at every stage, the verbal approximation cues had a one-to-one correspondence with the numeric metrics. This indicates that sequence in which information is presented is important. A choice of a sequential framework or hierarchical framework with sequential pipelines, lets you arrange information stages in specified manners within these pipelines. This would allow information to be presented to a human in the same way, albeit different form, as its presented to the statistical model being used in the analysis. Having such an arrangement, along with critical assessment of the information stages, should therefore, ensure, final conclusions and decisions made by a person, to be aligned and congruent with the results of the statistical model. This follows directly from the conclusions of Meder et al \citep{meder_sequential} and forms the basis for the structure of our interface framework.

Going back to our setup, we have statistical information coming at different stages of the analysis. As long as, at each stage, there is contemplation on the information by the examiner, the information at that stage will automatically be a part of the larger global cognitive strategy of decision-making that every individual has \citep{heck_information_2017}. These strategies are latent in nature and not observable. The only thing that we can observe is, the judgement and decision made at the end. A deeper discussion of these strategies is out of scope of this paper, but an important point that lies here is that, it has been shown that these strategies exist \citep{heck_information_2017} \citep{broder_newell_2008}.

These latent strategies affect the eventual decision-making, even though its very hard to distinguish one cognitive strategy from the other when the only point you have, is the final decision. \citep{broder_newell_2008} \citep{glockner_2009}.

Therefore, if we were to have calibrated trust in the statistical model, our framework should enable a global cognitive decision-making strategy for the examiner, that is in close correspondence with the steps followed by the statistical analysis. This would fall in lines with the findings of Meder and Mayrhofer \citep{meder_sequential}. A hierarchical structure that bridges the gap between domain knowledge and statistical results should allow for some associations to be drawn between the raw image data and the different kind of information from a statistical analysis. This is true because of two reasons. First, the visual comparison of the two objects (bullets) under a comparison microscope, also provides a psychological strategy and rational for making decisions. Second, as long as we give the right framework to the examiners, which has elements of raw data in it, the framework should serve as a bridge and a way to form implicit correspondence between the cognitive decision-strategies of the visual comparison task and the statistical analysis pipeline.

In their work, Heck et al \citep{heck_information_2017} talk about how people integrate multiple sources of information to draw inference in probabilistic settings. Information is processed sequentially by a person and, recommendation or heuristic for the probabilistic result is available for all sources of information. They first present a probabilistic version of the decision-strategy, take-the-best (TTB) heuristic, which is a heuristic based on ranked order of error probabilities given for the sequential processing of information. Then they show that between different cognitive decision-strategies adopted to make a decision, only 4 out of 104 people use the decision strategy take-the-best heuristic. Everyone else tried to integrate all available information in some kind of weighted-additive strategy to approximate the rational decision. \citep{heck_information_2017}

\begin{figure*}[htbp]
\includegraphics[width=\columnwidth]{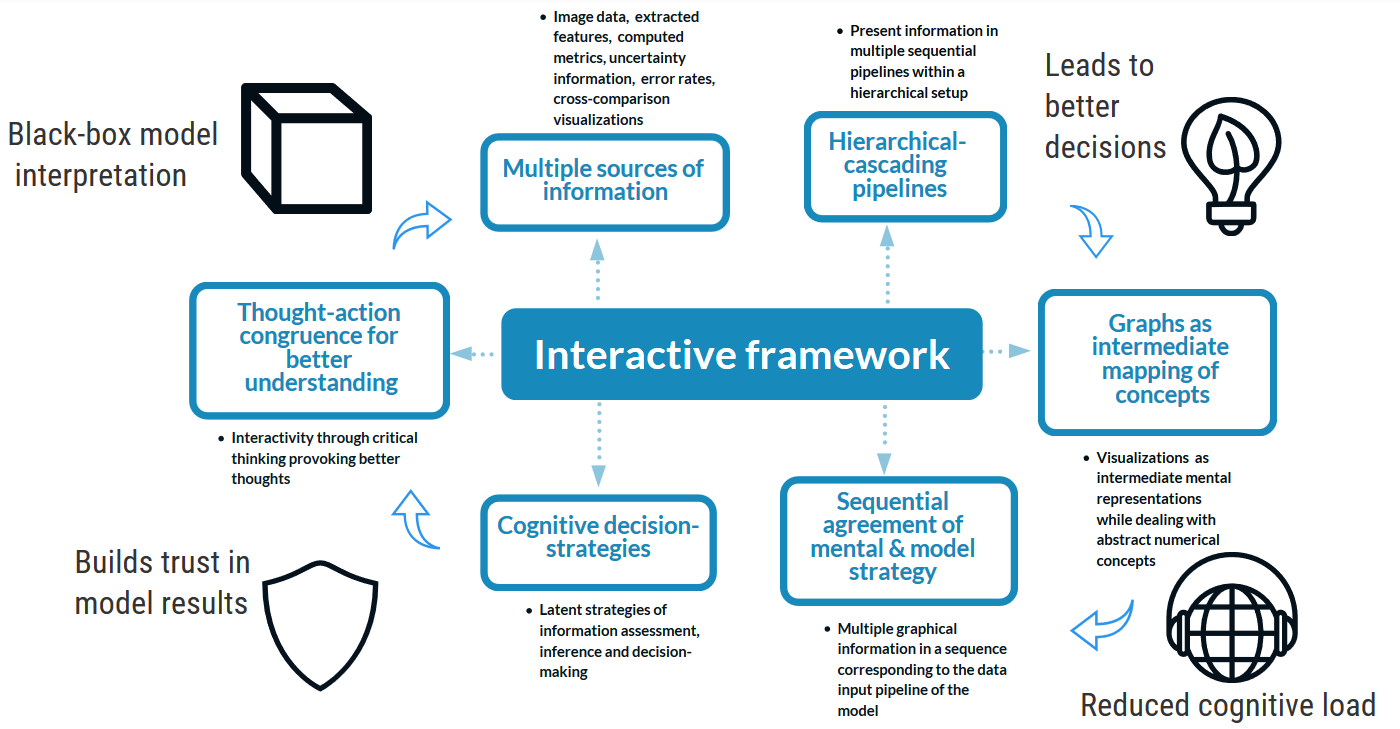} \caption{Concepts in cognition that connect to form the proposed cognitive principles of design for an interactive framework. The interactive framework ensure better decisions by calibrating trust in the model, reduction in cognitive load and enables interpretation of black-box results}\label{fig:cognitive_model}
\end{figure*}

What this means for our framework is that, providing recommendations on what a particular probabilistic metric or result means or what is the recommended decision at a particular stage, is not the best course of action. This is true especially when there is multiple sources of information coming in at different stages of analysis. This is because people are more likely to try to integrate all information that they encounter at different stages in some way or the other.

Another problem with providing straight on recommendations, is, it does not promote thought provoking actions that are investigative in nature. It hampers critical reasoning which might lead to either complete disregard of the recommendation or lead to blind trust. Both these things take away the transparency achieved by critical reasoning and logical decision-making and again brings subjectivity into the picture.

Therefore, we need to ensure by design, that the structure of the interface presents information in a manner, that would lead to the correct inference, while relying on critical thinking rather than recommendations. A hierarchical setup has information in specific pipelines which can complement the sequence of the steps of analysis. This follows the sequential approach used by Meder and Mayrhofer \citep{meder_sequential} and therefore cognitive decision-making strategies adhering to this setup should lead to better understanding of the results, thereby leading to better trust in the model.

A point that has to be duly noted here is that sequential presentation of information in the form of visualizations does not mean a sequential spatial arrangement of visualizations is necessary. It means that the visualizations, metrics and new information should be added sequentially in a temporal manner which conforms to the setup of the underlying statistical model. Interactivity allows this willful addition of stages. Every stage of information added, will therefore, be considered and contemplated upon before a new stage of information or visualization is added. This ensures consideration of every stage to some level of detail, without a straight away neglect and moving on to new source of information.
The mere requirement of using, for example a check-box can ensure that a willful decision of adding a new visualization has been made.

Following the suggestions by Heck et al.~\citep{heck_information_2017}, this interactive and sequential addition of stages and contemplation of each stage, holds importance in the larger scheme of decision-making, as people try to integrate all information in some way or the other.

For the case of firearm examination and bullet matching, we saw a few examples of visualizations that exist. The likes of LEA scans, cross-cut locations on the LEA, profiles, signatures, visualizations with extremas like peaks and valley, automatically aligned signatures etc can all serve as stages of information as they all bring some kind of new knowledge and abstraction. These can be used as bits and pieces in a cognitive decision strategy to connect to the model results. The key remains to do it in a manner that leads to accurate inferences, calibrates trust and motivates critical assessment. This is where the structure of cognitive design as suggested up till now comes in, to ensure, that first what the model is doing can be understood, and secondly all information is taken into account.

For an objective analysis, where now we have data in the form of metrics available at different stages, the latent decision making strategy, is a pipeline comprising of information gathered from these stages. This analysis is different from the visual comparison task. The visual comparison task will take all the data (the whole image of a bullet) into account. Any latent cognitive decision-making strategy employed, uses partial information from within the raw-data i.e.~images. Another important point with visual inspection is that, the whole image is made available in the beginning of this subjective analysis, and is sufficient to form internal strategies of comparison. Other external information presented like some metric is usually not included or sparingly included in visual assessment. This means the examiner has internalized understanding of key aspects for the visual assessment many of which are not quite explainable.

Now with this in mind we again consider the objective analysis. Here, we have new metrics and visualizations etc that the examiners can consider one after the other and keep moving to the next one. In this way, the examiners would have contemplated about a given metric at a given stage for a sufficient amount of time, before getting new information from a new stage of the analysis. This means each stage definitely has a role to play in the latent cognitive decision-making strategy.

Till now we have discussed the rational behind having a hierarchical framework and its importance in ensuring that the decisions made by the examiner after interacting with it, is compliant with the decision and recommendation suggested finally by the model scores. Note that by the recommendation of the model, we mean the take away point that a statistical or model expert would infer from the final score. This should not be confused with within the framework and stage specific recommendations that was discussed earlier.

This said, we go a little deeper into the principles that allow us to create a framework that justifies calibrating trust in the model through critical reasoning, thereby enabling better decision-making while keeping the cognitive load to a minimum. For this we again follow principles that are grounded in user studies.

Zhou et al \citep{zhou2017} in their work, described the effect of uncertainty and cognitive load on user trust. They used results from a machine learning or black-box model and monitored the performance of users on prediction of a prescribed quantity. The uncertainty information was presented in the form of graphs for different cognitive levels. The users were asked to rate their trust in the system when being tasked with drawing inference from the graphs. The performance on interpretability and predictive decision making of the users was recorded for machine learning experts, non-machine learning experts (researchers in other fields) and non-experts (people who neither had understanding of model nor were researchers). They concluded that the presence of uncertainty information leads to increased trust in the system but only under low cognitive loads. Presentation of uncertainty under high cognitive load leads to decreased trust in the model.

First, their work gives clear indication of how black-box or less interpretable model based results, can lead to different levels of trust in the system. Second, under conditions of low cognitive load, visualizations of uncertainty lead to increase trust in the model and better overall decisions based on the black-box model results.
Therefore as long as the cognitive load levels are kept low, providing visualizations of the uncertainty is better.

In lines with this, Allen et al.~\citep{allen2014} in their work, examined the effect of cognitive load on decision-making when uncertainty is communicated through graphs. The authors tested the ability to accurately derive information from 5 different graphs where probabilistic outcomes were defined. They concluded that cognitive load manipulation did not have an overall effect on derivation of information from the graphs (mean and probability information), but did suppress the ability to optimize subsequent actions that was based on this derived information and uncertainty around it. They concluded that cognitive load affected the use of complicated graphs. They also concluded under limited cognitive resources, the derivation of more complex information from graphs is hampered.

Both these work give support to the theory that as long as cognitive load is kept low derivation of information is accurate from graphs, calibrates trust in the underlying model and leaves room for making correct inference from the information. This necessitates our framework to keep the cognitive load as low as possible. In order to ensure this, we propose interactivity in the hierarchical framework proposed earlier. The interactivity ensures a sequential consideration of information. This means when we are presented with multiple stages of information like different scores, graphs, visualization, we only consider one of them at a time, then proceed to add a new stage after sufficient amount of time has been spent contemplating the current information. The need for this arises because most of the information gathered from graphs may not be simple inferences. Therefore, when we have multiple types of graphs, scores, tables etc, trying to make parallel considerations, it takes up a lot of cognitive resource. This is equivalent to the high cognitive load tasks that Allen et al.~\citep{allen2014} discuss in their paper. Keeping the consideration of information sequential and on a need-to-add basis, enables cognitive resources to be dedicated to one stage of information at a time. Given the critical nature of forensic examination, consideration of all such stages of information coming from the statistical analysis is necessary for critical reasoning and sound decisions. A sequential array in a hierarchical framework adding stages one at a time, is the simplest way to ensure that the cognitive load remains at a minimum, given that consideration of all information is mandatory.

This said, the argument of having more visualizations to convey abstract information was also evaluated by Gattis et al.~\citep{gattis}. These authors investigated the impact of mapping conceptual information to spatial relations. They proposed when needing to draw inferences, graphs provide some level of intermediate mental representations which enables people to move from visuospatial representations to abstraction. This is because people are able to use natural mental mapping between perceptual and conceptual relations better.

These conclusions from Gattis et al\citep{gattis} ensure that accurate inference is drawn from each stage of information in the sequential and interactive hierarchical setup that we have proposed. It is because visualizations and graphics provide the intermediate mental representation needed for better inference of abstract information. This also makes sure that at least from a design perspective, we have provided a means for accurate inference at each stage of information. And each accurate inference, contributes to a coherent latent cognitive decision-making strategy

Segal et al.~\citep{segal}, in their work, show how conceptually congruent actions can promote thought. Here, the authors investigate if action can support thought. They examine whether actions that are conceptually synchronized with the thought, can lead to other thoughts. They also investigate whether a one-to-one synchronization between thought and action facilitates better performance. To which, they conclude that direct actions, which are congruent with thoughts, leads to better performance than any indirect action. They concluded through their experiment that thought is an internalized action and re-externalizing thought through congruent mapping of actions can lead to other thoughts. In other words, congruence of thought and action supports understanding.

The evidence given by Segal et al\citep{segal} in support of congruence between action and thoughts support understanding and more thoughts, provides a premise for critical reasoning through investigative actions. This is the theme for our framework. The interactive nature of visualizations and the hierarchical setup for accessing different stages of information being produced by the statistical analysis, allows synchronized interaction with the framework. Within the framework, stages of information are provided one at a time, in a sequential and hierarchical setup. This means there is enough time and room for contemplation and thought to gain some understanding at each stage. This thought provoking step allows for critical assessment of the derived information from scores, metrics and visualization at the stage under consideration. The critical assessment then provokes an investigative action, which is usually some form of interaction with the framework. The investigative interaction then allows the examiner to add a new stage of information in the form of some new result or metric or visualization. The choice of the new stage of information is a direct result of the previous step of critical assessment. Therefore, the new addition in the sequential pipeline of information, is a product of congruent thoughts and actions. As this was initiated by a critical assessment, the new stage of information should provoke better thoughts. This sequence of actions and critical thinking is embodied in the interactivity of the framework, and ensures that the latent cognitive decision-making strategy leads to better understanding and more trust in the model.

These cognitive principle give us a solid foundation for the design of our framework. Therefore, we provide access to model results at different levels to encourage firearm examiners, to engage both with their own data, and the results from the statistical model. We want to strengthen the connection between data and the model. For that, we envision a user interface that works on separate layers. We wish is to present information in a manner that lets the firearm examiners investigate the process from the top to bottom through various kinds of interactions. Through the principles of cognitive theory as presented earlier, the interface reduces uncertainty and leads to better understanding before the examiners make a decision. The synchronization of interactions and information at different levels, should at the very end of the experience with the interface, let the domain experts make more confident decisions. The cognitive load introduced by evaluating uncertainty in results and scores from these complex predictive black-box models should decrease, as the investigative interactions of the domain expert increases. The investigative interactions in a congruent framework should lead to more confident decisions.

\hypertarget{critical-reflection-through-a-realization-of-the-cognitive-framework}{%
\section*{Critical reflection through a realization of the cognitive framework}\label{critical-reflection-through-a-realization-of-the-cognitive-framework}}
\addcontentsline{toc}{section}{Critical reflection through a realization of the cognitive framework}

\begin{figure*}[htbp]
\includegraphics[width=\columnwidth]{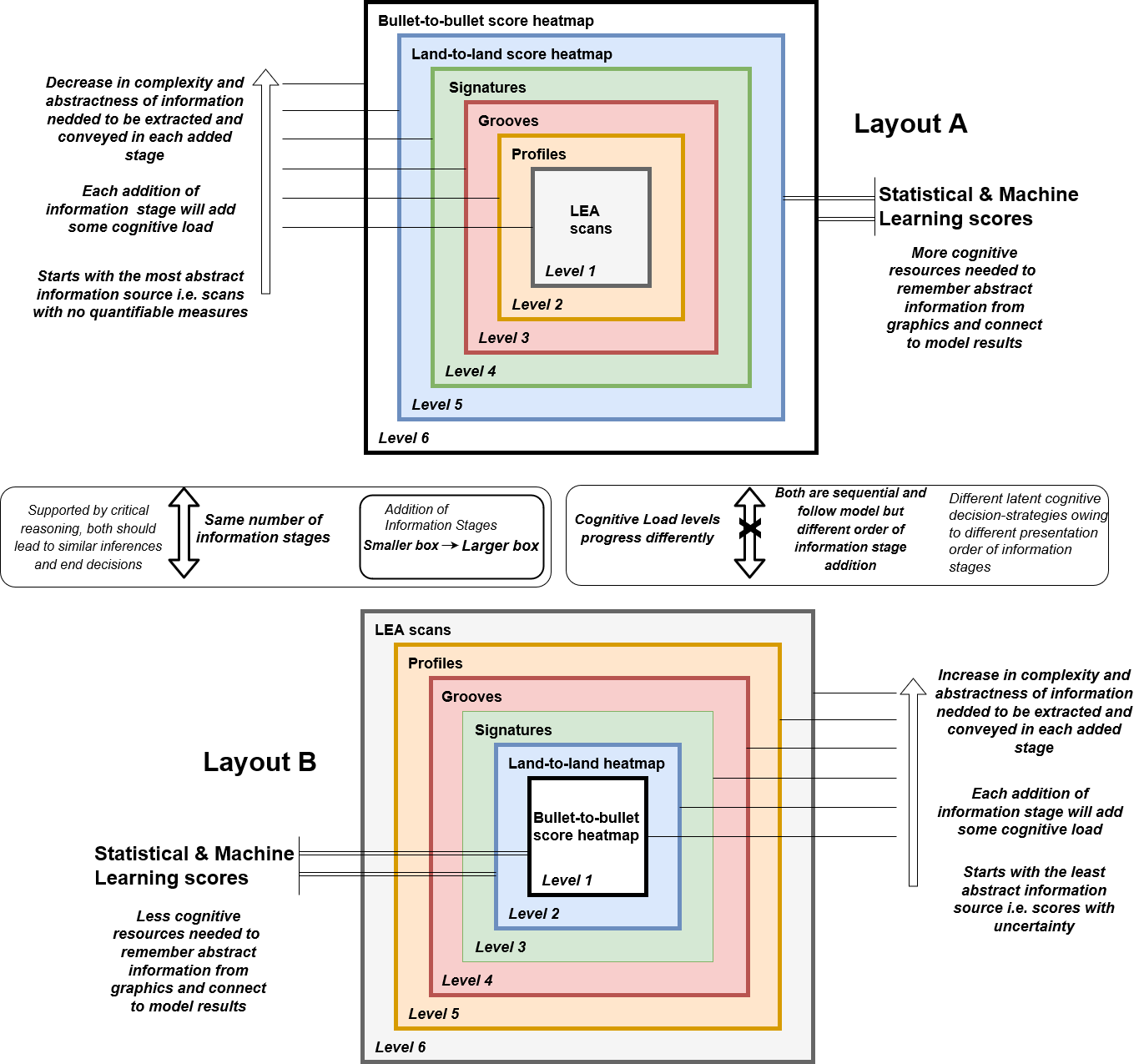} \caption{ The figure shows two possible layouts of a hierarchical decision ensemble with different levels of information. The first stage is automatically presented and contemplated upon and some level of abstraction of information is done mentally. Each newly added information stage is contemplated upon and the derivation of information adds to the derived information from previous stages, subsequently being a part of a latent cognitive decision strategy. Each information stage becomes a decision stage as soon as there is some contemplation, critical assessment and derivation of information from it. And because each new information is adding to the cognitive load from the previous stage, all sequentially added stages after due contemplation, will encapsulate the information from the previous stage. The concentric boxes represent this encapsulation of derived information. The largest box is the last added stage, after which again a critical assessment should lead to drawing of final inferences and decisions.}\label{fig:information_decision_ensemble}
\end{figure*}

For deriving the right conclusions in firearm examination, the framework we propose would be in the form of a hierarchy of contemplation stages where the stages represent information from the analysis level. The sequential framework of stages means it leads to a critical assessment of information at that level by connecting to abstractions made at all or any of the previous stages. Each hierarchical contemplation stage in the pipeline is a chart, visualization, score or raw data which is a direct outcome from some level of the objective statistical analysis that the examiner investigates. These are so chosen, that they allow the examiner to map results from the objective analysis to the images of the surface of the bullet. Any addition in information, can be viewed as an addition of a contemplation stage. Therefore, the latent cognitive decision flow through these contemplation stages leads to a final decision when all the stages have been added and contemplated upon.

Figure \ref{fig:information_decision_ensemble} shows two such possibilities of layouts, in context with objective firearm examinations. These two layouts can be seen as hierarchical ensembles of information abstraction which can lead to similar latent cognitive decision stages, owing to the sequential nature and following of the steps of the statistical and machine learning modeling procedure. As discussed earlier, the sequential nature of both these layouts in congruence with the modeling steps, should ensure that similar inferences are drawn in the end, as long as cognitive load balancing is not considered a problem. This is because the outermost and largest box in both layout A and layout B in the figure \ref{fig:information_decision_ensemble} represents the final stage that was added. In a latent cognitive decision strategy that is employed by a human, and in this case the examiner, it means after the final stage is added, the two layouts will have exactly the same information available at the disposal of the examiner, which leads to information equivalence in both. As long as the investigative critical reasoning behind both the layouts is able to connect the dots logically and correctly, the order of information should not have an affect on the final inference drawn regarding the comparison of the pair of bullets at hand. In reality, ensuring that correct inferences are drawn from the two layouts is not trivial or observable. But what can be ensured is, that, we choose the layout that corresponds to the cognitive theory of framework design in all aspects and not only the sequential representation part. This ensures the experience leads to better trust in the model for examiners, critical assessment of the results and accurate drawing of inference. In order to make this choice, its important to understand what sets the two apart and why one of them comes out as an obvious realization of the cognitive theory of framework presented earlier in all aspects.

Apart from the sequential and model congruent steps common in both the layouts, there are some key differences in the two layouts that set them apart. The cognitive load at each stage of the layout is one such factor that comes into play, and is very different in both these layouts. Since cognitive load has a very important role to play, as shown in the cognitive framework theory, when making critical decisions like those in forensic and judicial settings, its effect cannot be neglected. Layout A starts with the information presented as the raw data at its level 1. The raw data or scans is not presented with any numbers or other insightful measures, which means the inference drawn from it does not have any succinct summaries that are explainable or realizable objectively by different people. This leads to the extraction of information from the raw data being highly subjective and abstract in nature. This is the same kind of situation that the current firearm visual examination procedures face, where explainability is hard because of the the abstract and unclear nature of the inferences drawn. Therefore, any inference that ends up being drawn from this step will be quite vivid and will have significant cognitive load associated with it. Even if we assume that the examiners are capable enough to remember this abstract inference, by the time they get to the model results, the cognitive load becomes too high to be able to dedicate enough resources for a critical assessment of the two different levels of model scores that are presented with uncertainty. Therefore, there is a big chance that the examiner will be more self-confident about the decision that they already have made in level 1, making it harder to sell the results of the objective analysis to them.

Moreover, in regards with self-confidence and trust in automated results, Muir et al \citep{muir_1994}\citep{muir_1996} showed that, a human with higher confidence in themselves than the trust in the automated system, would override automated results. This means that there is need to make sure that the results of the model are taken into consideration first. In layout A, when the LEA scans are presented at level 1, the examiner would be more likely to reject the model results presented in the end, because they would have already formed an opinion on the land-to-land comparisons at hand based on their visual comparison. This would defeat the purpose of having the examiners go through the framework, and would in all likeliness lead them to override the model results. This, as we can see, is direct conclusion from Muir's work \citep{muir_1994}\citep{muir_1996}

Layout B, on the other hand does not have any of these problems, as it guarantees correct drawing of inferences through cognitive load balancing at key steps of information abstraction while calibrating trust in the system. Thus, layout A only partially follows the principles of the cognitive framework while layout B follows all the principles of the cognitive framework presented in the previous section.

To understand this better its important to examine the layout B as shown in Figure \ref{fig:information_decision_ensemble}. We can see that this instance of layout B has 6 stages of information in it. The first stage or level 1 has a heatmap of the bullet-to-bullet scores. The bullet-to-bullet scores, as seen in Figure \ref{fig:bullet_to_bullet}, shows a grid of comparison scores. These are the highest level of model results. Since this is level 1, it means it is presented first. The reason for starting with model results comes from the cognitive framework theory. It was argued by Zhou et al.~\citep{zhou2017} that, increased trust in the model results presented with uncertainty, can happen only under low cognitive loads, this was especially true for non-model experts. Therefore, the statistical model and machine learning scores are presented as levels 1 and 2 respectively. This ensures that contemplation and abstraction of information from these model results, happens when the cognitive load is as minimum as possible. Critical assessment and relative comparison of scores can then be done and accurate statistical inference can be drawn from these scores, in light of the error rates of the model and uncertainty introduced by the presence of many comparison scores.

Since level 1 is presented first, and level 2 is presented second; the assessment, decisions and inferences drawn from level 1 will be the cognitive load, as it will be kept in mind, when the level 2 is added. The heatmap visualization of level 1 does not vanish and remains accessible, but after inferences are drawn, it remains as representation for reference purpose. For critical reasoning, some level of abstraction will always remain in mind, after the contemplation on the level has occurred. This is because we are dealing with a connected system where all the information that has been or will be presented, has one purpose, that is to enable the firearm examiner to draw inferences from the model and connect it to what they understand about bullet comparisons. The addition of the level 2 is interactive in nature and can be added after the examiner is satisfied to an extent, with what they have inferred from level 1. This means the interactive action of adding the new stage is congruent with the mental investigative need of the examiner. The congruent investigative actions through interactivity was argued by segal et al \citep{segal} to promote better thoughts. Sub-(level 1) interactions like hovering over the heatmap to get comparison names, meta-information and numeric scores are also possible, and remain a product of the ``congruence of thought and action to promote better thoughts'' concept introduced by Segal et al.~\citep{segal}. Moving on from this, the addition of level 2 will mean a new level of model scores and uncertainty. Here again the examiner contemplates and thinks on what the 6 in-phase land-to-land scores mean in the light of other land-to-land comparisons within the selected bullet-to-bullet comparison. Figure \ref{fig:land_to_land} shows the land-to-land random forest score heatmap. We can see that the diagonal shows the scores for the 6 in-phase pairs of LEAs from a selected bullet-to-bullet comparison from level 1. The new information of level 2 also has a connection to the level 1 and therefore, any inferences or decisions drawn from level 1 are carried forward as a cognitive load.

These inferences/decisions of level 1, 2, 3, etc can be thought as mental encapsulations of inferences, which in simpler terms can be thought as the associated cognitive load of each level or stage. Each of the lower ``level'' or stage remains nested, within the higher ``level'' or stage that has been added. The higher ``level'' or stage will therefore encapsulate, the inferences drawn from that stage and all previous stages.

This can be imagined as concentric boxes, with the inner most box representing level 1 and the outermost box representing level 6. The cognitive load can be thought as the space or volume occupied within these boxes, and at each level, the cognitive load, is the cognitive load from the previous level box, plus, some added cognitive load due to inferences drawn from the current level. The outermost box, level 6, being the last stage of information added, is the largest box. Therefore, the volume it would occupy, would be the volume or cognitive load of level 5 box plus a little bit more of volume or cognitive load from inferences drawn in level 6.

Therefore, while making final decisions, the cognitive load of level 6 is the load that the examiner has to work with, to make any final inferences, recommendations or decisions.

This nestedness as explained, comes because each addition of a level or stage, means information drawn in the current stage is done so in-light with the already available inference from the previous stage. Since not all the levels or stages are presented in one go, it guarantees that the cognitive load addition is as explained.

There is also a need to coherently establish the connections of the two levels of model results and scores, before moving on to making connections with the LEA scans. This is important because the next levels would lead to better understanding of why a score is high or low or seemingly erroneous.
In the figure \ref{fig:information_decision_ensemble}, this dependence of different levels of model results i.e level 1 on level 2 is shown by the concentric encapsulation of the level 1 box within the level 2 box. To be explicit, the information that can be derived in level 2 (land-to-land scores) adds to the inference of what has been derived in level 1 (bullet-to-bullet scores) and should lead to better understanding. The land-to-land scores for example tell a more detailed story whether all 6 in-phase pairs have similar scores or non-similar score, and how much of an effect it has on the level 1 bullet-to-bullet scores etc, and as stated, establishing this connection brings better understanding. Doing this in the beginning means, enough cognitive resources are available to first develop some investigative rationale, which can be explored through the addition of the other levels.

Moving forward, we first note, that the act of adding a stage from level 1 to level 2, immediately brought forward new scores and uncertainty that needs to be inferred correctly in light of the level 1 results. After due contemplation and critical reasoning, the formation of a latent cognitive strategy motivated toward understanding the scores better, begins. Now, the level 3 which has the aligned signatures, can be added again because of the investigative need for more information.

Addition of level 3, occurs by directly interacting with the level 2 heatmap of the land-to-land scores, and selecting the land-to-land comparison that needs to be investigated further.

Similarly, the groove identification information of level 4 and side-by-side profile comparisons of level 5 and finally the two LEA scans can all be added through checkboxes. The additions are all done after due contemplation as described earlier, ensuring that correct inferences can be drawn given in-light of the inferences of the previous levels.
Any mismatch in inference, should therefore, instigate a critical diagnostic reasoning to rationalize and bring to light some other things that the examiner probably missed out on.

Apart from this, there are a few other points from the cognitive theory presented earlier, that can be verified in the current realization of the framework.

For example in the cognitive theory presented and the work of Heck et al\citep{heck_information_2017}, it was argued that in order to promote critical-reasoning out and out recommendations are not good. The layout B and the choice of visualizations included, do not give any such recommendations, instead they are all representations extracted from some part of the bullet matching pipeline (figure \ref{fig:bullet-matching-pipeline}). Therefore, to draw accurate inferences from these graphs it is essential to employ critical reasoning that not only allows abstraction at the current stage, but as said before, does so in the light of the inferences drawn from other previous levels. This is a vital cog in the cognitive theory of framework presented earlier. None of the visualizations in layout B, namely, signature, profiles, grooves or model scores provide any out and out recommendations of how this should be interpreted. Considering Layout B, when the final stage or level 6 is reached, the final decision to be made about the comparison, will include inferences from all previous stages in some weighted kind of strategy. This is what Heck et al \citep{heck_information_2017} argue in their work.

In regards with load manipulation and cognitive loads, as shown in the cognitive theory of framework earlier, Allen et al \citep{allen2014} have argued that, higher cognitive loads can lead to suppressed ability in deriving of information from complicated graphs with uncertainty. This means if we want to calibrate trust in the model results, we need to present the model results in the beginning as done in layout B. As argued earlier, this leaves enough cognitive resources to first draw inferences from the model results, and because the model results are presented very early on, any new levels like signatures and profiles can be added with explainability of the model scores in mind. This means the level 3, 4, 5, and 6 visualizations are all there to support or diagnose the model results. Moreover, proper contemplation at each stage, and interactive addition of new stages only when instigated by the need for new information, means that the cognitive load should already be kept as small as possible within each level, even though in a global sense with new levels the cognitive load increases. This would hold because in-depth exploration with the intention of understanding and critical analysis at each level, should lead to some coherence. Moreover, given the nature of the framework, all the information is still visually available for re-investigation if necessary.

The calibration of trust in the user interface framework and layout B is also supported by some other works in literature. O'Donovan \citep{odonovan} in his work, showed that trust can be built overtime as long as the recommendation error rates associated with the system results are presented, which gives an understanding of how the system can be expected to perform. While, Lee and Moray concluded that when the situation involves two decision-makers or recommenders namely a human and a machine or system, trust in the system results, is associated with the joint performance of humans and the machine or underlying algorithm.

Therefore in order to calibrate trust in the model results, we need to ensure that both examiner's confidence in their own visual comparison results and their trust in the model results, both get adjusted together.

The layout B, allows for this to happen as the examiners first starts with the model results and sequentially moves from bullet-to-bullet scores to land-to-land scores and so on to the LEA scans which the examiners are familiar with. This means they are the last stage i.e level 6 of layout B. At this stage all the information has been presented and available to the examiner, while the examiner has already gone through various levels of critical thought process to understand what the model is doing. Therefore, as showed earlier, the visual assessment of LEA scans is being reinforced by the inferences drawn from all the previous stages. At this stage, the examiner does not make deductions from the LEA scan independent of their deductions and inferences from the model results. Their visual assessment inferences are all drawn, with the inferences from the model results kept in mind.

Given the subjective and non-quantifiable nature of the visual assessment, it would be hard to override the results of model for no reason. And if at all the examiner decides to override the results, many of the symptoms of the problem in the model results would have already presented themselves through other level of visualizations like the signatures, grooves, profiles etc, and they examiner would be bound to only override it given justifiable cause.

Because of the nature of the layout B, the trust in the model results, would be bound to be built as the joint understanding of self performance and the model performance.

From O'Donovan's conclusions, for our framework, it also means that we provide a means for the examiners to not only map their beliefs to the model results, but also provide an opportunity to intuitively assess their own performance with respect to the performance of the model. A provision for this is made in the level 2 (land-to-land scores) of layout B, but before presenting this, it is important to review another work that fall in line with this.

\begin{figure}
\includegraphics[width=0.5\textwidth]{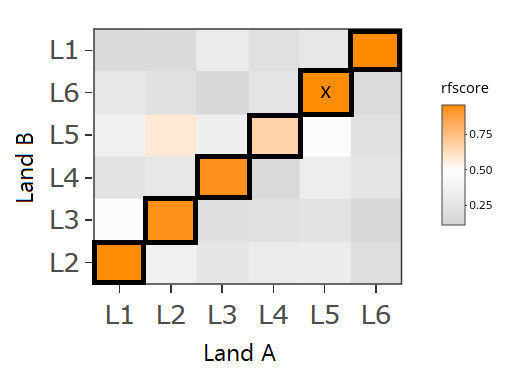}
\caption{The land-to-land random forest score with the "Match" option enabled. The black frame on the comparisons on the diagonal, shows 6 pairs of in-phase comparisons, that should have random forest score close to 1. We see that 5 of the pairwise comparisons have scores close to 1, but one of the pairwise comparison, Land 5 of Bullet B and Land 4 of Bullet A, has a score around 0.6. Adding this frame, shows that for an expected matching pair of bullets, all the 6 pairwise comparison on the diagonal with the black frame should have high score. Since this is not the case, the examiner, would need to go and investigate the signatures, grooves, cross-cut locations, signature alignment and LEA scans themselves to find, understand why this is happening. It might be because of problems in the automatic procedures of identification, or it might also be because the LEA is itself damaged for some reason. The change in score of one pairwise comparison, from the expected result, can also impact the bullet-to-bullet score which can be seen in figure 11  of the bullet-to-bullet score heatmap where Bullet A is B3 and Bullet B is B2}
\label{fig:match_land_to_land}
\end{figure}

Yu et al \citep{yu} in their work argue that, first it takes a few trials of interacting with an automated system to calibrate trust. Users are able to perceive and understand the results of system performance and results, and update their confidence in their own performance and adjust their trust and decision schemes accordingly.

Secondly, they showed that the trust in the system can grow and good perception of the system performance can be achieved along with better self-confidence in their abstraction methods, as long as the accuracy of the automated system results is more than 70\%.

Assessment of system performance for black-box models, is not a trivial task, especially when there are many complex steps involved in the underlying model. In order to bridge this gap between self-confidence in examiner's own assessment and model results, we present a feature in our framework to make a direct comparison of how the examiners self assessed conclusions stand when compared to the model results.

This feature is therefore presented as an option for the examiner through a check-box, on the land-to-land score heatmap (level 2 in the layout B). The feature essentially gives the examiner an option of selecting according to their belief if the selected bullet-to-bullet comparison is a ``Match''. This means if the examiner believes that the two bullets being compared, should be coming from the same gun, and a binary decision of ``Match'' is possible, then what are the land-to-land comparisons associated with it that should have scores close 1 (scores are on a 0 to 1 scale). Figure \ref{fig:match_land_to_land} shows the land-to-land comparison heatmap for a selected pair of bullets. The black boxes on the diagonal show, that if the examiner was expecting the bullets to be a match, then all the land-to-land comparisons with the black frame around it, should be close to 1 in the random forest scores. Therefore when the examiner applies this frame on the land-to-land heatmap, they can immediately see the difference in their belief of their self-performance and the performance of model results.

In the case that of the 6 pairs that are expected to have a near to 1 score, one or two comparisons show lower scores, the examiner can interact with the user-interface to add new visualizations like the aligned signatures (level 3 of layout B) to see if there is a problem there, or add level 4 and 5 of layout B, which shows the location of grooves and the quality of the extracted profile. This can tell the examiner, if the location of the grooves was incorrectly identified and thereby resulting in lower scores or maybe the extraction location of the cross-cut on the LEA scans was incorrectly chosen, or maybe it was something else.

In extension of the work of Yu et al \citep{yu} this should eventually lead to better understanding of the system performance and their own performance, which inturn would let them adjust their trust and latent decision-making strategies.

The validity of trust calibration through such an inferential framework can also be judged from prior research in neuroscience \citep{little_learning_2013}
\citep{gordon_toward_2011} \citep{bubeck_pure_2009} \citep{pfeifer_how_2007}. Little et al.\citep{little_learning_2013} explain that action-perception loops are the primary driving factor in learning and inference, and study exploration in the absence of external reward. They take the human agents learned internal model, and show that learning-driven exploration has an evolutionary advantage that lies in the generalized utility of an accurate internal model. They also demonstrate that agents that learn efficiently through exploration tasks, are able to better accomplish a range of goal-directed tasks.

This result shows how the congruent nature of the action-perception or more specifically action-thought and the internal model or latent cognitive-rationales, leads to better capture of the complex structure learning task of the data-to-score model pipeline presents and that the inferential framework simplifies. Moreover, as elicited by Little et al.\citep{little_learning_2013}, this leads to better accomplishment of a range of internalized goal-directed tasks. For forensic comparisons, these goals are accurate conceptualization of the scores and thereby, better communication of those results even in the presence of uncertainty.

Gordon et al.\citep{gordon_toward_2011} and Pfeifer et al.\citep{pfeifer_how_2007}on the other hand revealed how embodiment, opportunities and constraints placed on the action-perception loop of an agent can facilitate information processing and learning. They also revealed that action strategies garnered by interaction with an external environment can shape the information flow that an active agent is processing.

This again, clearly exemplifies that the inferential framework that we provide, can accommodate not just better learning, but also stratify the conceptual flow of information, by consistent interaction with the interactive system, which in turn constrains the inherent internal model by the opportunities for exploration and assessment provided through the framework.

\hypertarget{discussion}{%
\section*{Discussion}\label{discussion}}
\addcontentsline{toc}{section}{Discussion}

One important aspect, in calibrating trust is that the examiners have sufficient trials of working with the inferential framework. Yu et al \citep{yu} suggested this in their work. Apart from this, it is also important to present some model results to the examiners, where the end decision about the comparison at hand is known (but not revealed until after using the user interface). This lets the examiner experience the user interface system, a few times, draw inferences in lines with the cognitive theory of framework presented earlier, and then check what it means when compared to the ground truth of the the comparisons. This is again a kind of self-assessment about how the examiners are forming decision strategies, and making inferences and decisions, when exposed to model results through the user interface framework. The model assessment experience and comparison with ground truth would also help calibrate trust in the model results, which is in lines with the conclusions presented earlier from the work of Yu et al \citep{yu}, O'Donovan\citep{odonovan}, Lee and Moray \citep{lee} and Muir \citep{muir_1994}\citep{muir_1996}.

\begin{figure*}[htbp]
\includegraphics[width= \columnwidth]{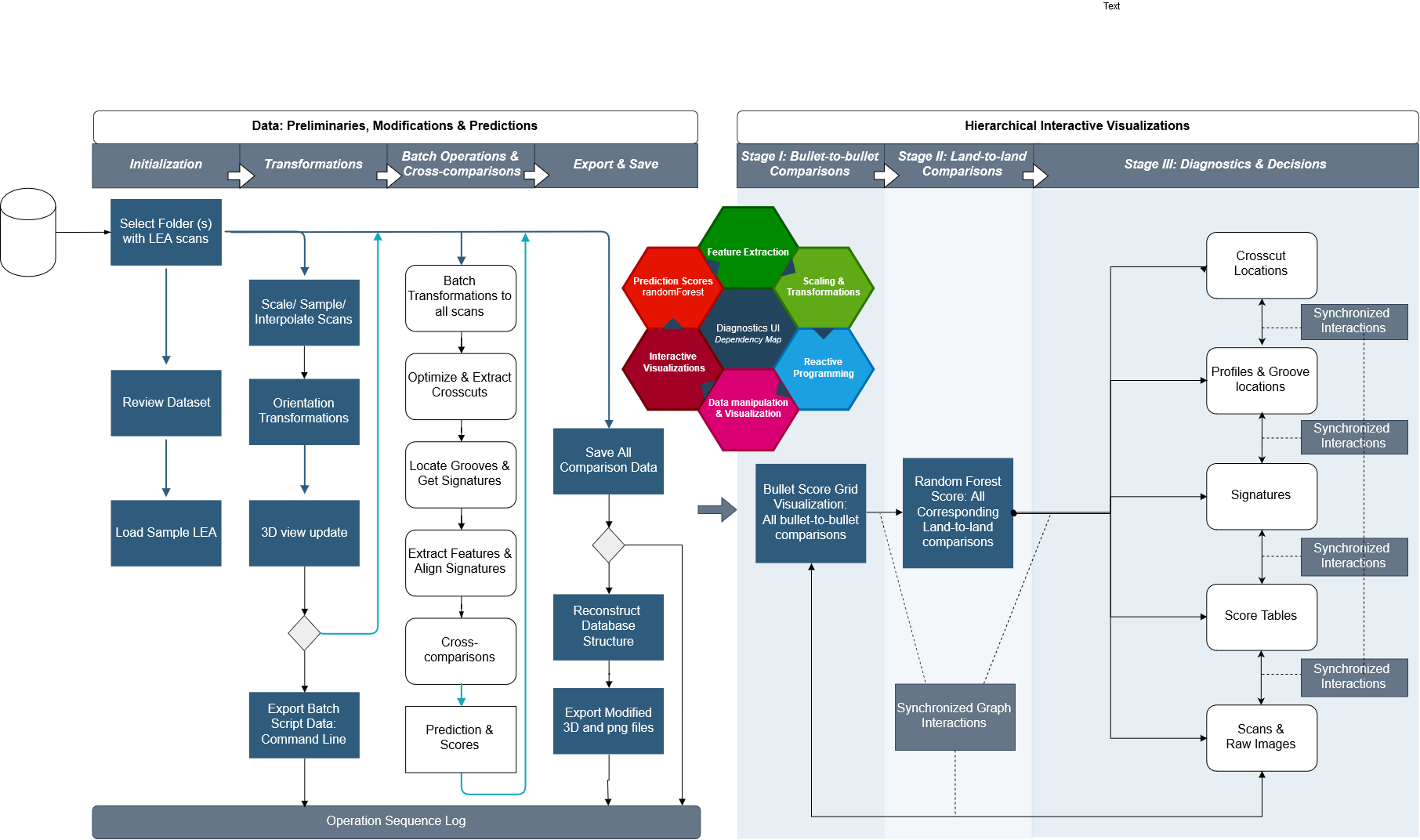} \caption{The flow chart shows the framework for an interactive diagnostics interface for the bullet matching pipeline}\label{fig:flowchart}
\end{figure*}

Once acceptable levels of trust have been established in the model results through the framework for the examiners, a more exploratory and diagnostic reasoning based approach can be suggested to them. This is important because it enables identification of model level problems or data level problems associated with the bullet matching pipeline. The bullet matching pipeline is shown in the figure \ref{fig:bullet-matching-pipeline} . The existence of multiple algorithmic steps that lead to the random forest model training means there can be multiple source of problems that can creep in from an algorithmic or model perspective. Therefore, all these different steps can be sources of problems which need to be investigated when the final random forest score seems suspect. This is already true when model experts are dealing with the data, but when domain experts deal with this, its even harder to diagnose the problems given their lack of understanding of whats gone wrong. Apart from the problems that are likely to occur within the pipeline, there can be other problems that can occur while making comparisons, that are inherent to the data.

Objective measures rely on the collection of high-resolution surface scans that captures information embedded on the surface of the bullet. The quality of the scans require close control. Standard procedures of scanning employ a controlled environment where the lighting, microscope table positions, zoom levels etc are kept constant. During the data collection, thousands of bullets for different firearms are collected. These scans serve as raw-data for objective measures of comparisons that are either trained on these scans, or use them for cross-validation. Depending on the firearm, some settings in the scanning procedure might be altered to ensure that the unique surface contours are captured correctly. Even with these standards in place, often times, there are problems that can creep into scans because of human error. The challenge here is that, it is not easy to identify whether we have a problematic scan, or whether the bullet itself was damaged or somehow uniquely different from the others. With hundreds and thousands of such scans, required to train machine learning models, it is also hard to manually verify the physical bullet with its scan at a later stage. Problems in the scans have a direct affect on model performance, accuracy and efficiency regardless of whether its a machine learning model or some other statistical procedure computing the similarity measure. But in machine learning which is inherently data drive, these problems can lead to serious issues. This is another kind of challenge which affects model results, and therefore diagnostic reasoning is of paramount importance in the bullet matching pipeline.

The interactive hierarchical visualization proposed in this paper, therefore, can not only help calibrate trust in the model results, it can do so while promoting diagnostic reasoning. Figure \ref{fig:flowchart} shows a clear flow of operations and actions on how an interactive framework can present results to the firearm examiners and allow them to diagnose the problems. A diagnostics mode is when the firearm examiners have some idea of where different pieces from the objective analysis fit in their cognitive decision strategy, that leads to congruent and accurate decisions between what the model result suggests and what the examiners conclude. This is important as it encourages critical assessment in special settings. From a model perspective, when the score is somewhere in between zero or one but away from these end points, and when there are no quantitative thresholds available, binary classifications cannot be made. Diagnostic reasoning is important in this situation. Critical diagnostic reasoning is required in many other situations too. For example, when the examiner is dealing with data where the ground truth in not known or when there is an obvious problem in the model results but the reasons are unknown or when it is not immediately evident why the model gives strong support for classifying the comparison one way over the other. It is also particularly important in the situations where the data is deemed unsuitable for any kind of visual comparison, but the model is still able to come up with a score. In the diagnostic mode, the information presented through interactions at different contemplation stages are driven primarily by critical assessment. Therefore, either more information can be investigated or different pipelines of information stages can be assessed simultaneously, if the examiner wishes to do so. This means a more holistic diagnosis of the comparison can be performed. The intent for the diagnostic mode is to enable the user to study and diagnose unaccounted problems in the data that might have led for an end-experience conclusion of ``unsuitable for decision making'' or ``insufficient evidence in data to make a decision of yes or no''. This mode is critical when dealing with real world scenarios, like freshly procured evidence from crime scene investigations.

The process of visual pattern recognition, which often relies on identification of certain data patterns and structure in the image, are usually intractable in terms of formal process elicitation and achieving objective robustness in the form of repeatability and reproducibility. We therefore, consider such data patterns as unidentifiable and latent parts of the more formal process of perceptual inference, which has been widely studied in neuroscience, cognitive psychology and visual inference \citep{gordon_toward_2011}\citep{little_learning_2013}\citep{heck_information_2017}\citep{buja_2009_statistical}. As such, we refrain from making claims about the exactness in assimilation of perceptually identified data patterns in the visual inference procedure, but focus more so on the systematic considerations inherent to human cognition and learning, that lead to visual or perceptual inference strategies.

By considering one sub-field of forensic examination as a proxy, we also attempt to stay away from specifics effects of perceptual pattern recognition that come from specific nature of data patterns and underlying structure, that in turn come from a specific class of images, or in this case specific forensic evidence. We therefore, focus on the general applicability of the framework to most sub-fields of forensic examination, without considerations for the effects of specific structures or the order of the perceptual consideration.

Thus, the descriptions and methods provided herein, although widely applicable to most sub-fields of forensic examination where visual comparisons are used, should be re-examined step-by-step when dealing with sub-fields other than forensic firearm examination.

\hypertarget{conclusions}{%
\section*{Conclusions}\label{conclusions}}
\addcontentsline{toc}{section}{Conclusions}

In this work, we have shown how to bridge the gap between domain knowledge of forensic examiners and AI/model results. We show that, in this transitional environment, when the field of forensics is shifting to more objective measures, there is a need for robust Human-AI collaboration systems. We show that there are frameworks that can be designed and implemented to address the problems faced by the forensic examiners. We show this in the context of firearm examination, that calibrating trust in the model and allowing correct inferences to be drawn from the results of objective models, are possible for domain experts. We show that critical assessment is key and visual explanations help map the domain specific understanding of the firearm examiner, to the model results. We ground our proposed framework by principles of cognitive theory. We show that having an inferential framework consequentially leads to calibration of trust. We then give an example within the context of firearm examination of how this framework can be realized as an interactive user interface. We also show a couple of examples through an application developed based on these principles, on how the tool can help understand different problems and connect the forensic evidence to scores, both statistically and perceptually.

\hypertarget{case-study-through-application}{%
\subsection*{Case study through application}\label{case-study-through-application}}
\addcontentsline{toc}{subsection}{Case study through application}

In this section we demonstrate the utility of the user interface for interactive exploration of results. The data set used here, was developed by FSI Houston. The bullets were chosen from a set of bullets used to test firearm examiners, and the barrels are Ruger LCP barrels. These barrels are differently manufactured when compared to the Ruger P85 barrels that were used to train the random forest. There are 5 bullets being compared here, out of which three (B1, B2 and B3) were fired from the same barrel, while the other two bullets are fired from unknown barrels, which means we do not know if they were fired from the same barrels as the bullets B1, B2 and B3.

\hypertarget{good-matches}{%
\subsection*{Good Matches}\label{good-matches}}
\addcontentsline{toc}{subsection}{Good Matches}

\begin{figure*}[htbp]
\includegraphics[width= \columnwidth]{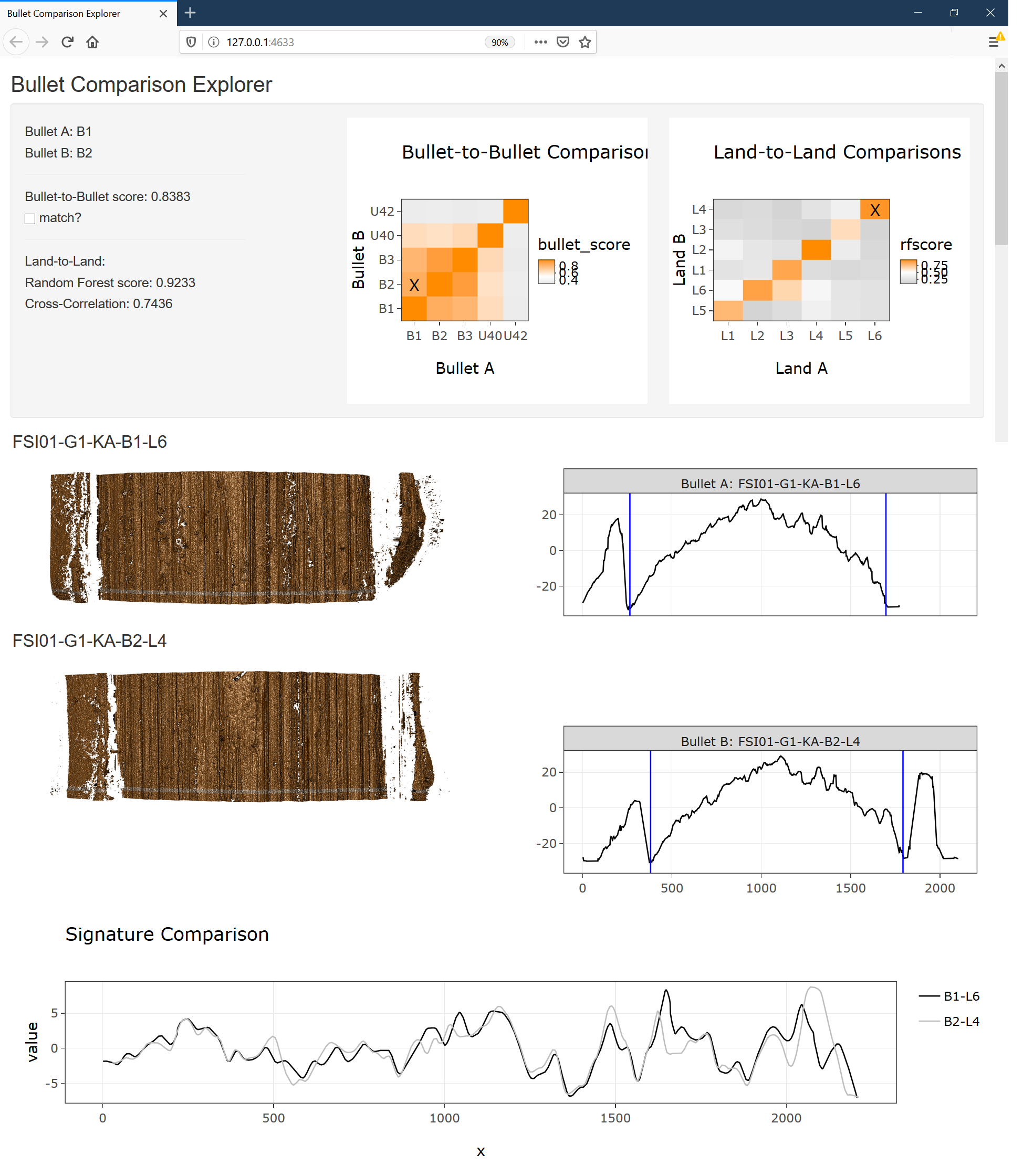} \caption{ Here we have an example of a good land-to-land match with a high random forest score of about 0.92. The profiles show a close correspondence for most of the part, but on the B1-L6 we can see that the right groove seems to be missing in the profile. The same can also be seen in the aligned signature correspondence. For most of the except the right most, the signatures seem to be well aligned with each other. This slight discrepancy might be the reason for a small reduction in the random forest scores. }\label{fig:goodmatch}
\end{figure*}

In order to understand how the application works, looking at a known match helps understand what does it mean when we say that two LEAs match. We start by selecting a particular comparison on the bullet-to-bullet level. Figure \ref{fig:goodmatch} shows the screenshot of the app after the last visualization has been added. The bullet B is chosen as B2 while the bullet A is chosen as B1, as we know that they were fired from the same barrel. We can see that the bullet-to-bullet scores for these two comparisons is on the higher side but not too near 1. This can help us realise even before looking at the land-to-land scores that maybe one or two of 6 pairwise comparisons do not have a high random forest score. Upon adding the level 2, i.e.~the land-to-land scores we select L4 for B2 and L6 for B1.

At this point if we had selected, the match check box, we would have got black frames similar to figure \ref{fig:match_land_to_land}, on the diagonal representing the expected land-to-land matches for the known matching bullet-to-bullet comparison. It is vital to realize that the feature merely seeks to present a means of understanding the difference between what we expect the scores to be and what they turn out to be. It does not seek to emphasize that binary decisions need to be made, but shows that when we know the bullets to be fired from the same barrel, there is a sequence of pairs that are expected to have high scores. This can be different from what happens in reality, when one or two lands of the bullet, due to the randomness of the firing process, might not have the same contact with the barrel of the gun as the other lands. This can very well mean that although ideally there should be 6 pairs of lands with high scores, in reality this is not going to be observed. This can also be a manufacturing characteristic, and specific to only this particular barrel of the gun. The main point behind this is that we cannot generalize that all 6 pairs of lands will have very high scores, but in an ideal situation, we would expect them to be that way. This is why it is important to understand that binary decisions are not necessarily the norm as even for known matches, there might be a few pairs of LEAs that can give non ideal scores.

Moving on to the example at hand, for the selected land-to-land comparison, if we go through adding different visualizations, in congruence with the interactive and cognitive framework rules as explained in the earlier sections, we will finally end up with the screenshot shown in the figure \ref{fig:goodmatch}. We see on the profiles that for B1-L6 the right groove is missing but on B2-L4 the right groove is present. Most of the part of the profile, 3/4 of it, seems in close correspondence for both the LEAs, but there is a slight change due to the absence of groove on the B1-L6. This effect can also be seen on the aligned signatures, where for most of the signatures, the alignment of the peaks and valleys seems to be good agreement with each other, except for the last 1/4 on the right. The land-to-land score is 0.92 which is pretty high, but the small dip in number might be due to groove related problem in one of the profiles which gets reflected through the various numeric inputs of the random forest model and finally in the random forest scores.

\hypertarget{unexpectedly-low-matches}{%
\subsection*{Unexpectedly low matches}\label{unexpectedly-low-matches}}
\addcontentsline{toc}{subsection}{Unexpectedly low matches}

\begin{figure*}[htbp]
\includegraphics[width= \columnwidth]{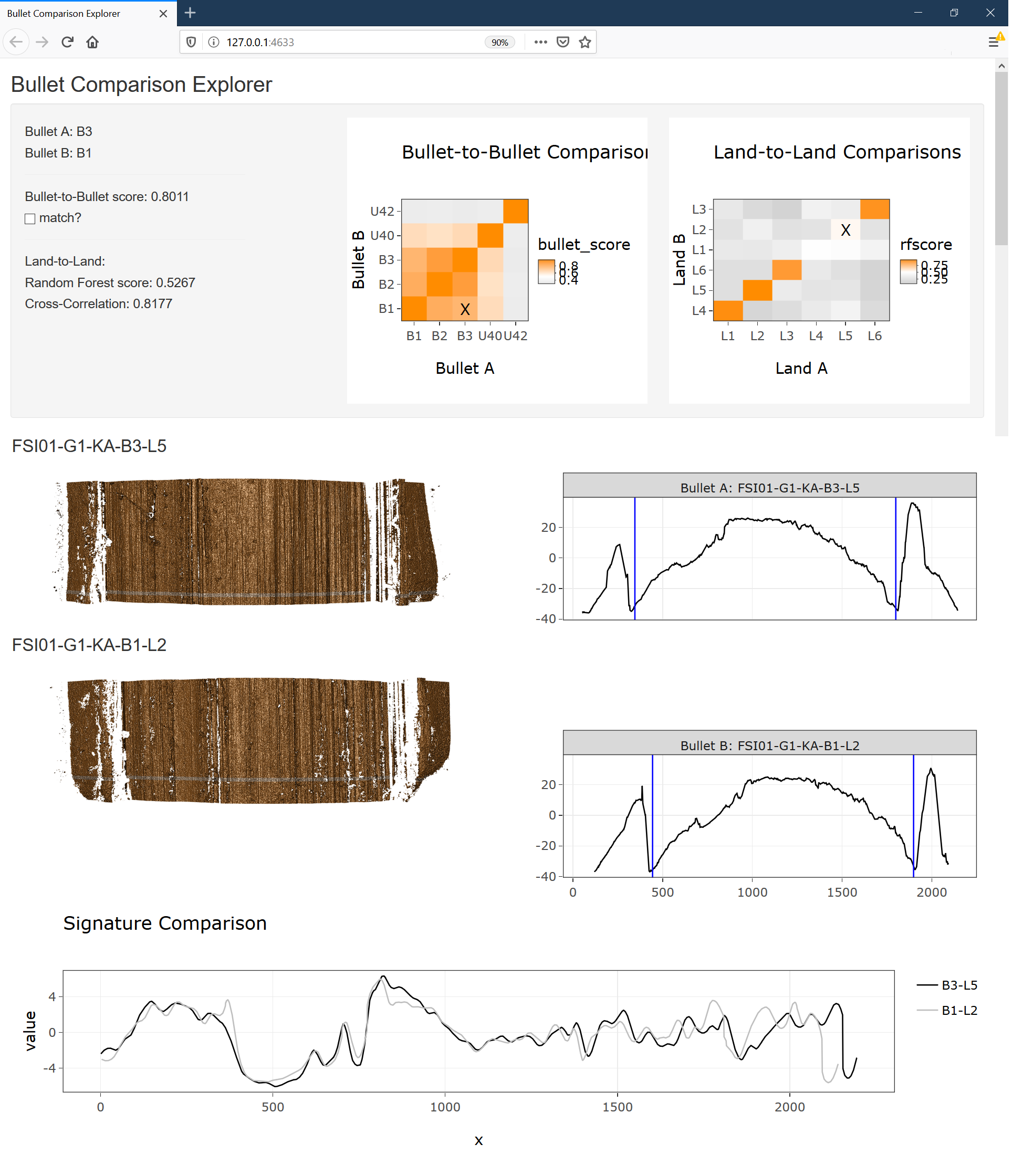} \caption{This example shows a bullet pair with a couple of land-to-land comparisons giving unusually low random forest scores. From the profiles it is hard to distinguish what exactly is the problem. The large scale structure seems similar at first, with correctly identified grooves. The aligned signatures on the other hand give a better picture as the right most 1/3 seems to not be aligned so well. On closer examination of the cross-cut locations, we can see that cross-cut extraction location, marked in the silver-grey horizontal line on the LEA scans, is at very different locations for the two scans. }\label{fig:notgoodmatch}
\end{figure*}

In order to examine how the diagnostic reasoning flows through the visualizations, we will examine a case where the bullets are supposed to be known matches but there are certain issues in the land-to-land random forest scores of a few pairwise comparisons. Figure \ref{fig:notgoodmatch} shows the screenshot of the app after the last visualization has been added. The application in figure \ref{fig:notgoodmatch} also shows some feature scores along with random forest and bullet-to-bullet scores on the left panel. An important point to understand here is, that at this stage it is assumed that the examiner already has interacted with the framework and has a deeper understanding of their trust in themselves and the model results at the same time. Given this, a diagnostic example where model results might be incorrect, is examined here. We start with choosing the Bullet A as B3 and Bullet B as B1. Since both these bullets come from the same barrel, all the in-phase land-to-land comparisons should have a high random forest score.
On selecting the bullet comparison, we notice that two of the 6 in-phase pairwise comparisons which are represented on the diagonal of the land-to-land comparison heatmap, have very low scores. We select the B3-L5 and B1-L2 as the pairwise comparison to investigate. The profiles, which are vertically scaled, seem to be similar in a first go, but on closer inspection we can see that there are some dissimilarities in the small scale structure that is hard to infer from. The grooves on both the ends of the profiles seem to be accurately identified, so those cannot be the problem. Then we go to the aligned signatures, and notice certain discrepancies. The rightmost 1/3 of the two signatures seem to be quite differently spread out and some other peaks around the left-middle do not seem to match very accurately. At this point, we move on to examine the LEA scans. Here we can notice that, the silver-grey horizontal line that represents the cross-cut extraction location, seems to be very different places for the two LEAs. For B3-L5 it seems to be closer to the base while for B1-L2 it seems quite a bit higher. This difference in extraction locations seems to contribute to having significantly different marking especially for B1-L2 where the extraction should have been near the base of the bullet. But when we look at the random forest scores and the cross-correlation scores, where CCF being largely associated with the aligned signatures visualization, we can see that there is a discrepancy. The CCF scores seem to be high enough, indicating that for this particular comparison, the random forest model seems to be choosing some numeric feature other than the CCF. From the reasoning it seems like there is a need to troubleshoot the model results more closely to investigate which numeric feature seems to be contributing to the RF scores and why. Therefore, when an adjusted level of trust based on critical reasoning is already established, a more diagnostic approach can be taken to figure out the cause for the unexpectedly low random forest scores for the current pairwise land-to-land comparison. This brings more understanding and clarity to how and why the model might be performing differently as compared what was expected by the examiner.

\bibliographystyle{ACM-Reference-Format}
\bibliography{bibligraphy-chi22.bib}

\end{document}